\documentstyle[onecolumn,psfig]{mn}
\begin{document}

\title[Tidal forcing of uniformly rotating star]
{Non-adiabatic tidal forcing of a massive, uniformly rotating star II:
The low frequency, inertial regime}

\author[G.J. Savonije and J.C. Papaloizou]
{G.J.~Savonije$^1$ and  J.C. Papaloizou$^2$\\
$^1$ Astronomical Institute `Anton Pannekoek', University of
Amsterdam and Centre for High Energy Astrophysics (CHEAF),\\ Kruislaan
403, 1098~SJ~Amsterdam, The Netherlands\\
$^2$ Astronomy Unit, School of Mathematical Sciences, Queen Mary \&
Westfield College, Mile End Road, London~E1~4NS}

\maketitle

\newcommand{\pdrv}[2]{\frac{\partial #1}{\partial #2}}
\newcommand{\drv}[2]{{{{\rm d} #1}\over {{\rm d} #2}}}
\newcommand{\ol}[1]{\overline{#1}}
\newcommand{\ul}[1]{\underline{#1}}
\newcommand{\eq}{\begin{equation}}
\newcommand{\ee}{\end{equation}}
\newcommand{\eqa}{\begin{eqnarray}}
\newcommand{\eea}{\end{eqnarray}}
\newcommand{\noi}{\noindent}
\newcommand{\nn}{\nonumber}
\newcommand{\vc}[1]{\bf #1}
\newcommand{\q}[1]{{\bf #1}}
\newcommand{\p}{\prime}
\newcommand{\bra}[1]{\left( #1 \right)}
\newcommand{\brac}[1]{\left[ #1 \right]}
\newcommand{\sigb}{\ol{\sigma}}
\newcommand{\omgb}{\ol{\omega}}
\newcommand{\ba}{\begin{array}}
\newcommand{\ea}{\end{array}}
\newcommand{\h}{\frac{1}{2}}
\newcommand{\hh}{\frac{3}{2}}
\newcommand{\iph}{i+\frac{1}{2}}
\newcommand{\jph}{j+\frac{1}{2}}
\newcommand{\jmh}{j-\frac{1}{2}}
\newcommand{\imh}{i-\frac{1}{2}}
\newcommand{\jphh}{j+\frac{3}{2}}
\newcommand{\Msun}{\mbox{${\rm M}_{\odot}$}}
\newcommand{\Lsun}{\mbox{${\rm L}_{\odot}$}}
\newcommand{\Rsun}{\mbox{${\rm R}_{\odot}$}}
\newcommand{\Omgs}{\mbox{$\Omega_{\rm s}$}}
\newcommand{\Omgc}{\mbox{$\Omega_{\rm c}$}}
\newcommand{\RS}{\mbox{$R_{{\rm s}}$}}
\newcommand{\Na}{N_{\rm \theta}}
\newcommand{\Nr}{N_{\rm r}}
\newcommand{\ZUN}{\mbox{\rm{i}}}
\newcommand{\RHM}{r_{\imh}}
\newcommand{\RHP}{r_{\iph}}
\newcommand{\DELH}{(\RHP-\RHM)}
\newcommand{\HP}[1]{{#1}_{\iph}}
\newcommand{\HM}[1]{{#1}_{\imh}}
\newcommand{\DELR}{(r_i-r_{i-1})}
\newcommand{\FUaI}{\left(\frac{P \chi_{\rho}}{r}\right)_{i}}
\newcommand{\FUbI}{\left(\frac{P \chi_{T}}{r}\right)_{i}}
\newcommand{\FUa}{\left(\frac{P \chi_{\rho}}{r}\right)_{\imh}}
\newcommand{\FUb}{\left(\frac{P \chi_{T}}{r}\right)_{\imh}}
\newcommand{\FNUHP}{\nu_{\iph}^{visc}}
\newcommand{\FNUHM}{\nu_{\imh}^{visc}}
\newcommand{\DPRI}{\frac{P_{\iph}-P_{\imh}}{\DELH}}
\newcommand{\DELRHI}{\frac{\HP{\rho}-\HM{\rho}}{\rho_i \DELH}}
\newcommand{\DPH}{\frac{\left(\frac{P_i-P_{i-1}}{\HM{P}}\right)} {\DELR}}
\newcommand{\droh}{\frac{\rho_i-\rho_{i-1}}{\HM{\rho}}}
\newcommand{\DROH}{\frac{\left(\droh\right)}{\DELR}}
\newcommand{\DFLH}{\frac{\left(\frac{F_i-F_{i-1}}{\HM{F}}\right)} {\DELR}}
\newcommand{\DTH}{\frac{\left(\frac{T_i-T_{i-1}}{\HM{T}}\right)} {\DELR}}
\newcommand{\ZVIS}{\ZUN \sigma \RS}
\newcommand{\ZH}{\ZUN \eta}
\newcommand{\delnu}{\mbox{$\Delta \nu$}}
\newcommand{\delmu}{\mbox{$\Delta \mu$}}
\newcommand{\xir}{\left(\frac{\xi_r}{\RS}\right)}
\newcommand{\fr}{\left(\frac{F^{\p}_r}{F} \right)}
\newcommand{\xit}{\left(\frac{\xi_{\theta}}{\RS}\right)}
\newcommand{\ft}{\left(\frac{F^{\p}_{\theta}}{F} \right)}
\newcommand{\xip}{\left(\frac{\xi_{\varphi}}{\RS}\right)}
\newcommand{\rh}{\left(\frac{\rho^{\p}}{\rho}\right)}
\newcommand{\Rh}{\frac{\rho^{\p}}{\rho}}
\newcommand{\tem}{\left(\frac{T^{\p}}{T}\right)}
\newcommand{\Tem}{\frac{T^{\p}}{T}}

\begin{abstract}

\noi We study the fully non-adiabatic tidal response of a uniformly rotating
unevolved 20$ \Msun$ star to the dominant $l=m=2$ component of the
companion's perturbing potential. This is done numerically with a 2D implicit
finite difference scheme.
We assume the star is rotating slowly with angular speed
$\Omgs\ll \Omgc$, so that the centrifugal force can be neglected, but the Coriolis
force is taken fully into account. We  study  the low frequency `inertial'
regime $|\sigb| < 2 \Omgs$, where $\sigb$ is the forcing frequency in the
frame rotating with the stellar spin rate $\Omgs$. In this frequency 
range inertial modes are excited in the convective core which can interact
with  rotationally modified g- or r-modes  in the radiative envelope  and  cause
significant strengthening of the tidal interaction.
Resonant interaction with quasi-toroidal (r-)modes in slightly super-synchronous
stars causes efficient spin down towards corotation.
We determine timescales for tidal spin-up and spin-down in the inertial frequency
regime for stars spinning with $\Omgs=0.1 \Omgc$ and 0.2 $\Omgc$.
\end{abstract}


\section{Introduction}

In a series of earlier papers we studied the tidal response of an early type 
star in a  close binary system (e.g. a massive X-ray binary),
ignoring stellar rotation, but taking into account the effects of internal
nuclear evolution (Savonije \& Papaloizou 1983, 1984: SP83,SP84 and Papaloizou
\& Savonije 1985: PS85). Recently, we extended these studies by taking into
account the effects of the Coriolis force in uniformly rotating stars (Savonije,
Papaloizou \& Alberts 1995: SPA95). We found that, as expected, in a rotating 
star resonances between the $l=2$ tide and internal modes are possible not only
for $l=2$, but also with modes with predominantly $l=4$ or $l=6$ etc. due to the
mode coupling by the Coriolis force. Once the relative forcing frequency $\sigb$
was smaller than twice the stellar rotation speed, however, the numerical 
results were swamped by short wavelength inertial waves. We could only study
the (positive frequency range of the) inertial regime by adopting the so called
`traditional approximation' in which the $\theta$-part of the stellar angular
speed $\Omgs$ is ignored. It appeared that rotationally modified g-modes
continued to be excited in this frequency range but that for $\sigb< \Omgs$,
no reliable results could be  obtained. Therefore a new implicit 2D scheme
was developed that is numerically more stable and that can treat
three-level difference equations, enabling viscosity to be introduced in the
equations of motion. The viscosity can then be used to damp the 
shortest wavelength inertial waves. 
We have used this new scheme to explore the inertial regime,
introducing a much finer grid resolution in the radial direction.
For free oscillations, the low frequency inertial regime was studied by e.g.
Berthomieu et al. (1978), Saio (1982) and by Lee \& Saio (1987, 1989).
However, these authors did not take the Coriolis force fully into account, but
used the `traditional approximation', or truncated (to the first few terms)
the infinite expansion in spherical harmonics required to describe the
oscillations in rotating stars. We take the Coriolis force fully into account,
by not using spherical harmonics expansions, but solving the full
2D problem in $r$ and $\theta$. Also we consider (tidally) forced oscillations.
The next section (2) summarizes our basic equations,  followed by a brief
section (3) on the stellar input model. Then, in section 4, we present our 
numerical results.
In the appendix we list the finite difference approximation to the partial
differential equations describing  non-radial oscillations that give stable
numerical results and a short description of the new implicit 2D elimination
scheme used to solve these equations.

\section{Basic equations}
We consider a uniformly rotating, early type secondary star with mass $M_{\rm
s}$ and radius $R_{\rm s}$ in a close binary with circular orbit with
angular velocity $\omega$ and orbital separation $D$.
We assume the secondary's angular velocity of
rotation $\vec{\Omgs}$ to be much smaller than its break-up speed, i.e.
$(\Omgs/\Omgc)^2\ll 1$, with $\Omgc^2=GM_{\rm s}/R_{\rm s}^3$, so that
effects of centrifugal distortion ($\propto \Omgs^2$) may be neglected in
first approximation. We wish to study the response of this uniformly rotating
star to a perturbing time-dependent tidal force. The Coriolis acceleration is
proportional to $\Omgs$ and we consider its effect on the tidally induced
motions in the star. We use spherical coordinates ($r,\theta,\varphi)$, with
origin at the secondary's centre, whereby $\theta=0 $ corresponds to its
rotation axis which we assume to be parallel to the orbital angular momentum
vector. We take the coordinates to be non-rotating.

As is well known, in a non-rotating star the solutions of the linearized
non-radial stellar oscillation equations can be expressed in terms of
spherical harmonics, i.e. the spatial part of each mode can be fully
separated into  $r$-, $\theta$- and $\varphi$-factors (e.g. Ledoux \& Walraven 
1958)
\[U(r,\theta,\varphi)=u(r) P^m_l(\cos \theta) {\rm e}^{{\rm i}m\varphi}, \]
where $P^m_l$ represents the associated Legendre polynomials for $l$
and $|m|$.

The introduction of the Coriolis force, however, destroys the full
separability of the oscillation equations, it only being retained for the
$\varphi$-dependence. It turns out (e.g. Berthomieu, Gonczi, Graff, Provost
\& Rocca 1978) that two
independent sets of approximately spheroidal oscillation modes exist: modes
in which the density perturbation is {\it even} with respect to reflection in
the equatorial plane, which have $l-|m|$ even valued, and modes with {\it
odd} symmetry for the density, having $l- |m|$ odd valued. In addition, for
each $l,$  there is a set of approximately toroidal modes (e.g. Papaloizou \&
Pringle 1978) which couple with the spheroidal modes of $l \pm 1$. 

Let us denote perturbed
Eulerian quantities like pressure $P^{\p}$, density $\rho^{\p}$, temperature
$T^{\p}$ and energy flux $\vc{F^{\p}}$ with a prime. The linearized
hydrodynamic equations governing the non-adiabatic response of the uniformly
rotating star to the perturbing potential $\Phi_T$ may then be written

\eq \left[ \left(\pdrv{ }{t} + \Omgs  \pdrv{ }{\varphi}\right)
{v}_i\right] {\bf e_{\it i}}+ 2 \Omgs {\bf k} \times {\bf v}^{\p}
=-\frac{1}{\rho}
\nabla P^{\p} + \frac{\rho^{\p}}{\rho^2} \nabla P - \nabla \Phi_T,
\label{eqmot} \ee

\eq  \left(\pdrv{ }{t} + \Omgs \pdrv{ }{\varphi}\right) \rho^{\p} +
\nabla.\left(\rho \vc{v}^{\p} \right) =0, \label{eqcont} \ee

\eq  \left(\pdrv{ }{t} + \Omgs \pdrv{ }{\varphi}\right) \brac{ S^{\p} +
\vc{v}^{\p}. \nabla S }=-\frac{1}{\rho T} \nabla.\vc{F^{\p}}, \label{eqe}
\ee

\eq \frac{\vc{F^{\p}}}{F}=\left(\drv{T}{r}\right)^{-1} \left[ \left(\frac{3
T^{\p}}{T} -\frac{\rho^{\p}}{\rho} -\frac{\kappa^{\p}}{\kappa} \right) \nabla
T + \nabla T^{\p} \right] .\label{eqf} \ee 

\noi where  ${\bf e}_i$ are the unit
vectors of our spherical coordinate system, {\bf k} is the unit
vector along the rotation axis, $\vc{v}^{\p}$ denotes the
velocity perturbation, $\kappa$ the
opacity of stellar material and $S$ its specific entropy. These perturbation
equations represent, respectively, conservation of momentum, conservation of
mass and conservation of energy, while the last equation describes the
radiative diffusion of the perturbed energy flux. For simplicity we adopt the
Cowling (1941) approximation, i.e. we neglect perturbations to the gravitational
potential caused by the secondary's distortion. We also neglect perturbations
of the nuclear energy sources and of convection. 

The dominant tidal term of the
primary's perturbing potential is given by the real part of

\eq \Phi_T(r,\theta,\varphi,t)= f r^2 \, P^2_2 (\mu) \,e^{2 {\rm i} (\omega
t - \varphi)} \label{eqpot} \ee

\noi where $\mu=\cos\theta$, $P^2_2(\mu)$ is the associated Legendre
polynomial for $l=|m|=2$ and $ f= -\frac{G M_{\rm p}}{4 D^3},$ with $M_{\rm
p}$ the companion's mass.
Adopting the same azimuthal $m=2$ symmetry
and time dependence for the perturbed quantities  as the forcing potential,
the perturbed velocity ${\bf v'}= 2 {\rm i}(\omega-\Omgs) {\bf \xi}$, where
$\vc{\xi}$ is the displacement vector. The perturbations can be written as e.g.

\[ \xi_r(r,\theta,\varphi,t)=\widehat{\xi_r}(r,\theta)\,{\rm
e}^{2{\rm i}(\omega t-\varphi)} \] where $\xi_r$ is the radial component
of the displacement vector and $\widehat{\xi_r}(r,\theta)$ is assumed 
complex to describe the phase shift $\Delta \varphi$ with respect to the
forcing potential (\ref{eqpot}).

For numerical reasons it proved better not to eliminate $\xi_{\varphi}$ from the
equations, as was done in SPA95, and retain the $\varphi$- equation of motion.
However, we do eliminate $F^{\p}_{\varphi}$ with help of the
$\varphi$-component of the radiative transport equation.
Writing for simplicity from now on $\xi_r$ for
$\widehat{\xi_r}(r,\theta)$ etc., while dividing out the factor ${\rm e}^{2{\rm
i}(\omega t-\varphi)}$ , equations (\ref{eqmot})--(\ref{eqf}) transform into
the equations (\ref{eqmr})--(\ref{eqFt}) given below. 

\noi The radial equation of motion becomes
\eqa \rho \sigb^2 \xi_r  
+ \left[2 \ZUN\sigb \Omgs \sin{\theta} \rho\right] \xi_{\varphi} 
- \pdrv{ }{r} \bra{P \chi_{\rho} \Rh} 
- \pdrv{ }{r} \bra{P \chi_T \Tem }
+\drv{P}{r} \rh  
- 6 f \rho r \bra{\sin{\theta}}^2 =0
\label{eqmr}
\eea

\noi where $\sigb=2 \omgb\equiv 2(\omega - \Omgs)$ is the forcing 
frequency felt by a mass element in the uniformly rotating secondary, while
$\chi_{\rho}=\pdrv{}{\ln \rho} \ln P $ and $\chi_{T}=\pdrv{}{\ln T} \ln P$
follow from the equation of state.

\noi The $\theta$-equation of motion becomes
\eqa
\rho \sigb^2 \xi_{\theta} 
+ \left[2 {\rm{i}}\sigb \Omgs \cos{\theta} \rho\right]  \xi_{\varphi} 
- \bra{\frac{P \chi_{\rho}}{r}} \pdrv{ }{\theta} \rh + 
- \bra{\frac{P \chi_T}{r}} \pdrv{ }{\theta} \tem  
- 6 f \rho r \sin{\theta} \cos{\theta} =0
\label{eqmt}
\eea

\noi The $\varphi$-equation of motion becomes
\eqa
\rho \sigb^2 \xi_{\varphi} 
-\left[ 2 \rm{i} \sigb \Omgs \sin{\theta} \rho\right] \xi_r  
- \left[2 \rm{i} \sigb \Omgs \cos{\theta} \rho\right] \xi_{\theta} 
+ \left[\frac{\ZUN m P \chi_{\rho}}{r \sin{\theta}}\right] \rh 
+ \left[\frac{\rm{i} m P \chi_{T}}{r \sin{\theta}}\right]  \tem  
+ 3 \ZUN m f  \rho r \sin{\theta} =0
\label{eqmp}
\eea

\noi The perturbed equation of continuity is transformed into
\eqa \frac{\rho^{\p}}{\rho}+\frac{1}{r^2\rho}\pdrv{ }{r}\bra{r^2\rho\xi_r} 
+\frac{1}{r \sin{\theta}} \pdrv{ }{\theta} \bra{\sin{\theta} \xi_{\theta}} 
- \frac{\rm{i} m}{r \sin{\theta}} \xi_{\varphi} =0 
\label{eqco}
\eea

\noi By applying the thermodynamic relation
\[ \delta S=S^{\p}+{\bf \xi}.\nabla S= \frac{P}{\rho T}\frac{1}{\Gamma_3-1}
\left(\frac{\delta P}{P} -\Gamma_1 \frac{\delta \rho}{\rho} \right) \]
where the symbol $\delta$ denotes a Lagrangian perturbation and $\Gamma_j$
the adiabatic exponents of Chandrasekhar, the perturbed energy equation can
be  transformed into
\eqa
\left[\drv{\ln P}{r}-\Gamma_1 \drv{\ln \rho}{r}\right] \xi_r  
+  \left[\chi_{\rho} - \Gamma_1 \right]\rh 
+ \left[\chi_T + {\rm i} \eta \bra{\frac{m}{r \sin \theta}}^2 
\bra{\drv{\ln T}{r}}^{-1}\right] \tem   +  \nn \\
-\ZUN \eta \left[ \frac{1}{r^2}\pdrv{}{r} \bra{r^2\frac{F^{\p}_r}{F}} +
\drv{\ln F}{r} \bra{\frac{F^{\p}_r}{F}} \right] + 
+\ZUN \eta \left[\frac{\sin \theta}{r}\pdrv{}{\mu}\bra{\frac{F^{\p}_{\theta}}{F}}
-\frac{\cos \theta}{r \sin \theta} \bra{\frac{F^{\p}_{\theta}}{F}} \right] = 0
\label{eqen}
\eea

\noi where \[ \eta=\bra{\Gamma_3-1}\frac{F}{\sigb P} \] is a local
characteristic radiative diffusion length in the star, with $F$ the
unperturbed (radial) energy flux. When $\eta \rightarrow 0$ the response
becomes adiabatic and shows (outside resonance) no phase shift with respect
to the companion. However, even for `high' frequencies $\sigb /\Omgc \approx
1$ the diffusion length $\eta$ becomes comparable to the scale height when
the stellar surface is approached, so that the imaginary part of the surface
response becomes comparable to the real part.

The perturbed radial energy flux is given by
\eq \bra{\frac{F^{\p}_r}{F}} 
=\bra{\drv{\ln T}{r}}^{-1} \pdrv{}{r} \tem + - \bra{\kappa_T -4} \tem
-\bra{\kappa_{\rho} +1}  \rh
\label{eqFr}
\ee

where $\kappa_{\rho}=\pdrv{}{\ln \rho}\ln\kappa$ and $\kappa_{T}=\pdrv{}{\ln
T}\ln\kappa$. Finally the $\theta$-component of the perturbed flux follows
as
\eq \bra{\frac{F^{\p}_{\theta}}{F}} =- \frac{\sin\theta}{r}
\bra{\drv{\ln T}{r}}^{-1} \pdrv{}{\mu} \bra{\frac{T^{\p}}{T}}. \label{eqFt}
\ee
\noi where $\mu=\cos{\theta}$.

\subsection{Boundary conditions}
The differential equations are supplemented by the following boundary
conditions: at the stellar centre we require $\xi_r$ and $F_r^{\p}$ to vanish,
while at the stellar surface, we require the Lagrangian pressure
perturbations to vanish

\eq \frac{P^{\p}}{P}+ \drv{\ln P}{r} \xi_r=0 \label{bsc1} \ee
\noi and the temperature and flux perturbations to fulfill 
Stefan--Boltzmann's law  \eq \frac{F^{\p}_r}{F}=\left(\frac{2}{r}+4 \drv{\ln
T}{r}\right) \xi_r + 4 \bra{\frac{T^{\p}}{T}}. \label{bsc2} \ee

Furthermore,  $\xi_{\theta}$ and  $F^{\p}_{\theta}$ must vanish on the rotation
axis while in view of the symmetry of the tidal force  we adopt mirror symmetry 
about the equatorial plane, i.e., for $\theta=\pi/2$ we also require $\xi_{\theta}$ and
$F^{\p}_{\theta}$ to vanish.

\subsection{Viscous damping of short wavelength response in convective 
regions}
In order to damp the short wavelength (of order gridsize) response
to the tidal forcing in
the inertial regime in the non-stratified (convective) regions of the star,
viscous terms are added to the equations of motion. 
We use a simple
mixing length prescription for both energy transport and turbulent viscosity
in convective regions.
We adopt a (constant) reduction factor $\epsilon$ for the turbulent viscosity
to roughly take into account the mismatch between  the forcing time and
the convective timescale in the core.
The turbulent viscosity $\zeta$ can thus be expressed as
\eq
\zeta=\epsilon \,\,\, H_P\left(\frac{F}{10 \rho}\right)^{\frac{1}{3}}
\ee
where $F$ is the local energy flux and $H_P$ the local pressure scale height.
For the turbulent viscosity in the convective core
a reduction factor  $\epsilon=10^{-3}$ is adopted (adequate to damp the
grid oscillations for all applied forcing frequencies),
while in the convective outer shell a slightly 
larger value $\epsilon= 3 \times 10^{-2}$ had to be  taken.
At the boundary of the convective regions we let the viscosity
$\zeta$ drop to negligible values  as
$\exp{\left[-\left(\frac{2d}{H_P} \right)^2\right]}$,
where $d$ is the distance from the boundary. Fig.~\ref{fIM}
shows the adopted $\zeta$ distribution through the star. A significant steeper
decline to zero viscosity at the convective core boundary would not prevent
the short wavelength inertial response to penetrate into the weakly
stratified regions adjacent to the core. Adopting a much steeper decline (decline
on scale  $0.1 H_P$) changed the tidal torque by less than a few percent.
In the convective surface shell near $i\simeq 1600$, the ratio of the forcing time
to the local viscous time $\tau_v=\frac{\Delta r^2}{\zeta}$ (where $\Delta r$ is
the local gridsize) becomes
larger than unity. But since the mass in these layers is negligible this
has no effect on the tidal torque, which is predominantly applied in the radiative
region adjacent to the convective core.

With this prescription for the viscosity in the convective regions
the grid oscillations are adequately damped while the global numerical results,
like the tidal torque, appear insensitive to the adopted viscosity.
Only if $\zeta$ is taken some orders of magnitude larger do the global numerical
results like the net tidal torque change significantly.

The (simplified) viscous terms that are added to the equations of motion are
expressed in such a way that the  PDE's (\ref{eqmr})-(\ref{eqmt}) retain 
their separability in $r$ and $\theta$ when $\Omgs \rightarrow~0$.

The radial equation of motion (\ref{eqmr}) is extended with the following 
viscous terms
\eqa
\ZUN \sigb \frac{\rho \zeta}{r^2} \left[ \left(1-\mu^2\right) \pdrv{^2}{\mu^2}
\xi_r -2 \mu \pdrv{}{\mu} \xi_r -\frac{4}{1-\mu^2} \xi_r \right] 
+\frac{\ZUN \sigb}{r^2} \pdrv{}{r}\left(\rho\zeta r^2 \pdrv{\xi_r}{r} 
\right)
\label{eqvisr}
\eea
\noi while to the $\theta$-equation of motion the following terms are added
\eqa
\ZUN \sigb \frac{\rho \zeta}{r^2} \left[\left(1-\mu^2 \right) \pdrv{^2}{\mu^2}
\xi_{\theta} -4 \mu \pdrv{} {\mu}\xi_{\theta} -\frac{5-2 \mu^2}{1- \mu^2}
\xi_{\theta} \right]
+\frac{\ZUN \sigb}{r^2} \pdrv{}{r} \left(\rho \zeta r^2 \pdrv{\xi_{\theta}}{r}
\right)
\label{eqvist}
\eea

\section{The unperturbed stellar model}
\begin{figure}
\psfig{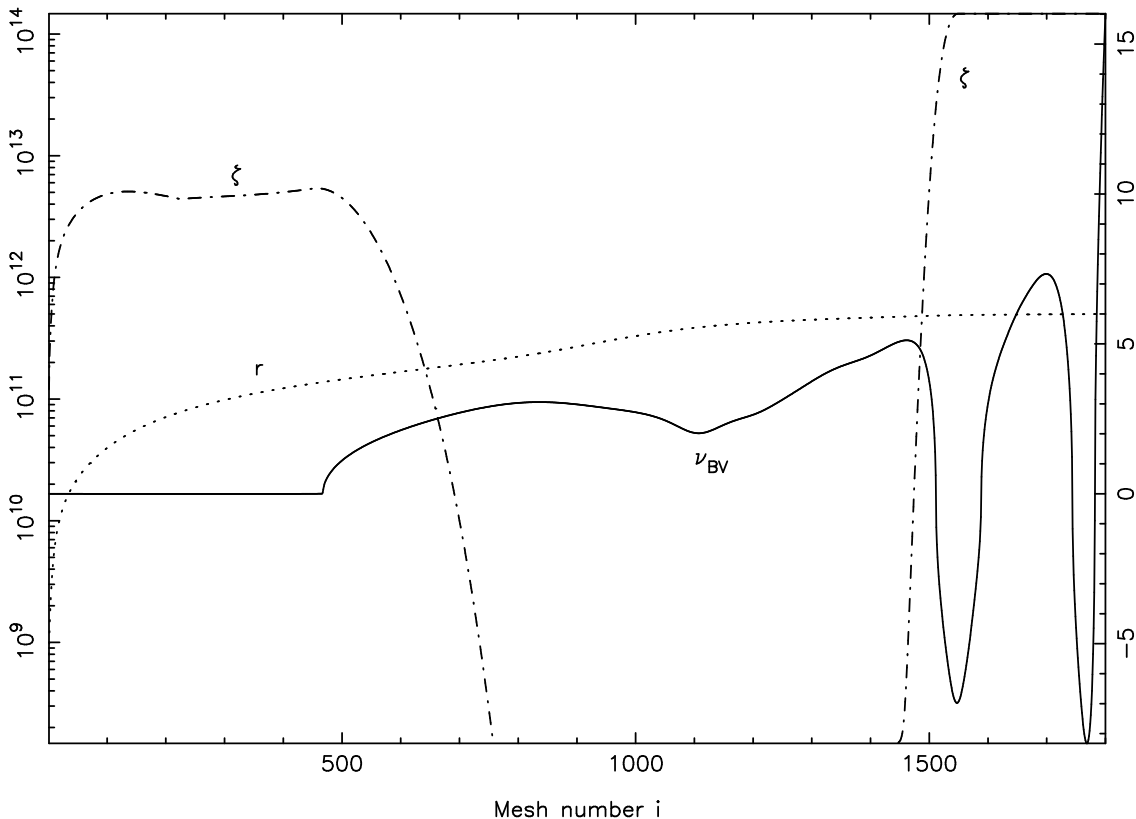}
\caption[]{Characteristics of the 20 $\Msun$ stellar input model:
the (adapted) Brunt$-$V\"{a}is\"{a}ll\"{a} frequency $\nu_{\rm BV}=$
sign${\cal A}\sqrt{|{\cal A}|}$ in units of the stellar break-up speed
$\Omgc$ (vertical scale on the right), radius $r$ in cm and turbulent
viscosity $\zeta$ in cgs units (vertical scale on the left) as a function
of radial mesh number.} 
\label{fIM}
\end{figure}

A recent version (Pols, Tout, Eggleton \& Han 1995) of  the stellar evolution 
code developed by Eggleton (1972) was used to
construct the unperturbed stellar input model for the secondary.
The model represents a spherical zero-age main-sequence star of 20 $\Msun$
with chemical composition $X=0.70$ and $Z=0.02$. The model was constructed 
with the OPAL opacities (Iglesias \& Rogers 1993). With these opacities
the 20 $\Msun$ ZAMS star appears to have  shallow convection zones in the
envelope which are absent when the Cox--Stewart opacities are used. 
The stellar radius  equals $R_{\rm s}=4.993 \times
10^{11}$ cm, while the stellar moment of inertia $I_{\rm s}=7.21\times 10^{56}$ g
cm$^2$. The break-up angular speed equals $\Omgc=1.46\times 10^{-4}$ s$^{-1}$.
In the convective core, which has a mass of about 10~$\Msun$, the imaginary
Brunt$-$V\"{a}is\"{a}ll\"{a} frequency  ${\cal A}^{\h}$ given by
\eq {\cal A}=
{1\over \rho}{{\rm d}P\over {\rm d}r}\left({1\over \rho}{{\rm d}\rho\over
{\rm d}r} -{1\over \gamma P}{{\rm d}P\over {\rm d}r}\right).\label{BVfreq}
\ee
\noi reaches a maximum
value of about $ |{\cal A}|^{\frac{1}{2}} =5 \times 10^{-3} \ll \Omgs$, where
$|{\cal A}|^{\frac{1}{2}}$ is given in units of $\Omgc$, while in the outer
convective shell $ |{\cal A}|^{\frac{1}{2}}\simeq 5 $, see Fig.~\ref{fIM}. 
This corresponds to a super-adiabaticity $\bigtriangledown -\bigtriangledown_
{ad} \simeq 10^{-6}$ and 0.07 in the core and envelope shell, respectively.

It can be seen that the radial grid has about 500 zones in the convective core,
about 600 zones in the adjacent radiative `plateau region' where
the eigenfunctions of g- and r-modes have most of their radial nodes,
and also several hundred zones in the rarified outer layers   with convective 
shells.

\subsection{Non-adiabatic surface layer (NASL) \label{sNASL}}
A crude random walk estimate for the time required for heat to diffuse to the 
stellar surface  from a given point $r$ in the stellar envelope  is given by
\[ \tau_{th}=\frac{\rho \kappa (\RS -r)^2 \beta}{c (1-\beta)} \]
where $\kappa$ is the opacity, $c$ the velocity of light, and $\beta$ the ratio
of gas to total pressure. By comparing this timescale with the forcing time
$\tau_f=2 \pi/\sigb $, we find that non-adiabatic effects become important
($\tau_{th} \simeq \tau_f$) roughly from
radial zone 1300 (for $\ol{\omega}=0.01 \Omgc$) or 1400 (for $\ol{\omega}=0.20 
\Omgc$) outwards.
The convective shells are thus located in this thin non-adiabatic layer, see 
Fig.~\ref{fIM}, so that the (adiabatic) Brunt$-$V\"{a}is\"{a}ll\"{a} frequency 
$|{\cal A}|^{\h}$ greatly overestimates the convective growth rate there
($\tau_{th}^{-1}\simeq 3\Omgc$ in the innermost convective shell). 
Also the thermal stratification in the radiative part of the non-adiabatic layer
is much weaker than suggested by the large values $|{\cal A}|^{\h} \simeq 5 \Omgc$
or even higher, attained in that region.

\section{Numerical results}
\subsection{Introduction}
We extend the calculations of SPA95 into the low-frequency inertial
regime, i.e. for forcing frequencies $|\sigb|< 2 \Omgs$. In SPA95 the numerical
results in this frequency regime were swamped by short wavelength inertial 
waves in the convective
core, which also propagated into the weakly stratified radiative regions adjacent
to the core, unless we adopted the `traditional approximation' in which the 
$\theta$ component of the stellar angular velocity $\vc{\Omgs}$ is neglected.
And even then, we could not obtain reliable results for $\omgb<0.1$, where
$\omgb=\h \sigb$ is the tidal pattern speed in the corotating frame
expressed in units of $\Omgc$.
From now on we shall express all frequencies  in units
of $\Omgc=(GM_s/\RS^3)^{\h} $.

Equations (\ref{eqmr})-(\ref{eqFt}) have been approximated by the set of finite
difference equations (FDE's) (\ref{fd1}-\ref{fd7}) given in the appendix.
The FDE's with appropriate boundary conditions are then solved by a newly
developed 2D implicit elimination scheme briefly explained in appendix B.
With this new three-level finite difference scheme we can effectively damp
the very short (gridsize) inertial waves by introducing a modest (turbulent)
viscosity $\zeta$ in the equations of motion, see equations (\ref{eqvisr})
-(\ref{eqvist}).

\subsection{Fourier-Legendre coefficients}
For each forcing frequency $\sigb$ the calculated stellar response can be 
expanded in Fourier-Legendre (FL)series with complex coefficients $C^m_l(r)$, e.g.

\[ \xi_r(r,\mu)=\sum_{l=0}^{\infty} C^m_l(r) P^m_l(\mu) \]
\noi with
\[ C^m_l(r)=\frac{(2l+1)(l-m)!}{2(l+m)!} \int^1_{-1} \xi_r(r,\mu) P^m_l(
\mu) {\rm d}\mu. \]

\noi By plotting the real and imaginary parts of $C^m_l(r)$ for the various
perturbed variables versus $r$ for $m=2$ and different $l$-values, we can compare 
the strength of the different $l$-components in the response.

\subsection{Tidal response in non-rotating stars}
To compare our results for (uniformly) rotating stars with those of non-rotating
stars we have recalculated (for $\omega<0.1$) the tidal response
for a non-rotating star with the same 2-D scheme and with the same viscosity. 
For forcing frequencies  in the range considered here the amplitude of the
displacement vector $\vc{\xi}$ globally follows that of the equilibrium tide,
except that in the subsurface layers ($i>1500$), a peak occurs which grows
with decreasing tidal frequency $\sigma=2 \omega$. For $\omega=0.025$ 
the radial displacement peaks at more than twice the value of the
equilibrium tide near $i=1500$, while $\xi_r$ decreases to the equilibrium value  near
the stellar surface. The $\theta$-displacement is generally an order of magnitude
larger than $\xi_r$  near the surface.
The mass in the NASL is only small, so that the effect on the total torque 
is generally insignificant.
The likely reason for this strong sub-surface response
is that for low forcing frequencies the radiative losses in the 
surface layers (section \ref{sNASL}) get so strong that the characteristic
timescale for
convective oscillatory motions becomes comparable to the forcing period.
Indeed, when a zero value for the radiative diffusion length $\eta$ (adiabatic
calculation) is adopted the  strong  sub-surface response disappears.

\subsection{Tidal response in sub-synchronous stars ($\Omgs<\omega$)}
For sub-synchronously rotating ($\omgb>0)$ early type stars rotationally modified
g-modes are excited  in the radiative envelope
for all frequencies in the inertial regime, down to near
co-rotation. The g-modes have a predominant $l=2$ component, with $l=4$ and 6
components having in general a factor 4-10 weaker amplitudes in this frequency
range.  If
the relative forcing frequency $\sigb=2 \omgb= 2(\omega-\Omgs)$ in the frame 
corotating with the secondary falls below $2 \Omgs$,  inertial waves are
excited  in the weakly or non-stratified layers. Once $\omgb<\Omgs$
the first signs of inertial waves  are observed, not only in the 
convective core, but even stronger in the non-adiabatic surface layer (NASL)
containing the convective shells (section \ref{sNASL}).
Without viscosity the tidal response in the convective regions gets overwhelmed by
short lengthscale (gridsize) variations, as discussed in SPA95.
The excitation of inertial modes in the core is discussed in section \ref{sinert}.

\begin{figure}
\psfig{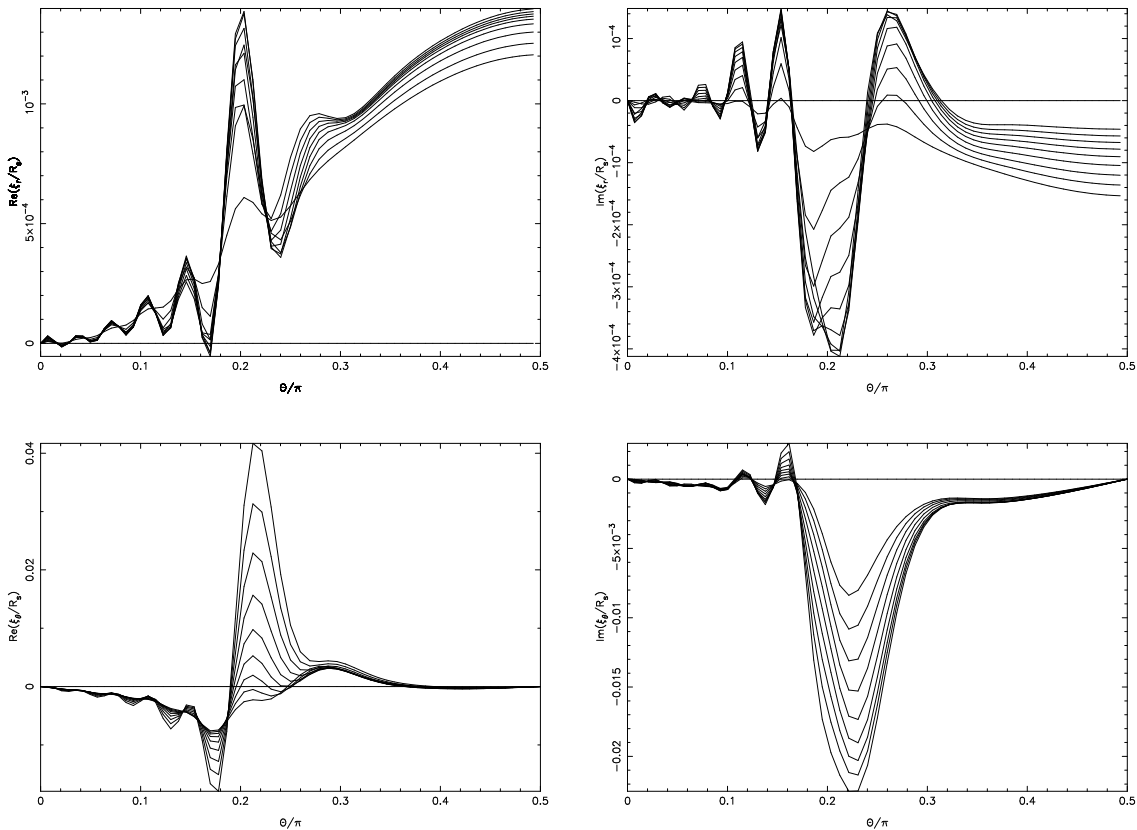}
\caption[]{$\omgb=0.076$ and $\Omgs=0.1$:
the $\theta$-dependence of the radial- and $\theta$-component of the
displacement vector for radial zones in the range $i=1650-1750$, corresponding to
$r/\RS=0.992-0.998$, i.e. in between the two convective shells near the surface.
The left hand panels give the real parts, the right hand
panels the imaginary parts. 
The shortest lengthscales were removed by viscous damping.}
\label{f076}
\end{figure}

As discussed above for non-rotating stars,
the strong radiative damping in the NASL gives rise to large amplitude oscillatory
convective motion in the sub-surface layers.
With rotation the non-adiabatic  response in the sub-surface layers shows
even larger $\theta$-displacements,  while the response is  more 
complex with high $l$-components due to inertial effects. 

Fig.~\ref{f076}  shows an example of the $\theta$-dependence of the tidal response
in the sub-surface layers of a star rotating with $\Omgs=0.1$ for a forcing
frequency $\omgb=0.076$.
At $i=1400$ (not shown) the
$\theta$ dependence is still typically that of the radiative region outside the
convective core and consists predominantly of a simple $l=2$ ($\xi_r$) or $l=3$ 
($\xi_{\theta}$) component.  The displacement $\xi_{\theta}/\RS$ attains a
maximum of about $8 \times 10^{-4}$ at $i=1400$. However,
slightly further out (from about $i=1500$)
inertial waves are excited which show up as short wavelength oscillations.
The amplitude of the displacement vector  $\vc{\xi}$ 
shows a peak at around $\cos\theta \simeq \omgb/\Omgs$ with an oscillatory decay
towards the rotation axis.  Note that g-modes become evanescent when the
oscillation frequency is smaller than twice the average radial component of the 
stellar angular velocity $2 \Omgs \cos{\theta}$ (e.g. Eckart 1960), 
so that the peak amplitude of the tidal response appears to coincide roughly with
the turning point of those modes. Inertial waves can propagate from pole to
equator.

\subsection{Tidal response in super-synchronous stars ($\Omgs > \omega$)}
When we lower the forcing frequency - through corotation to negative values -
the tidal response becomes qualitatively different. Now the excited tidal wave
becomes retrograde as the star is spinning faster than the orbital revolution
of its companion. For small $|\omgb|$ there are again many radial  nodes in the 
radiative envelope, as there are for small positive frequencies, but now the
displacements show a strong toroidal component. When $|\omgb|$ is increased a
rapid succession of strong resonances occurs, these are resonances with free
r-modes, e.g. Papaloizou and Pringle 1978, Provost, Berthomieu \& Rocca (1981),
Rocca (1982) and SPA95, 
analogous to Rossby-waves in the earth's atmosphere (e.g. Pedlosky 1977).
Because the excited r-modes occupy only a small frequency range
approximately given by
$ -\frac{m \Omgs}{l (l+1)} < \omgb < 0 $ with $l=3$ and $m=2$, strong resonances
with low order (with few radial nodes in the envelope) r-modes
are possible for relatively small $|\omgb|$, i.e. close to corotation.

Hence, in contrast to early type {\it sub}-synchronous stars approaching 
corotation, which tend to spin up towards corotation as a result of weak
resonances with high-order, strongly damped g-modes, slightly
{\it super}-synchronous stars tend to spin down towards corotation  much more
efficiently due to strong resonances with weakly damped r-modes
(see Figs.~\ref{ft001}-\ref{ft002}).

\begin{figure}
\psfig{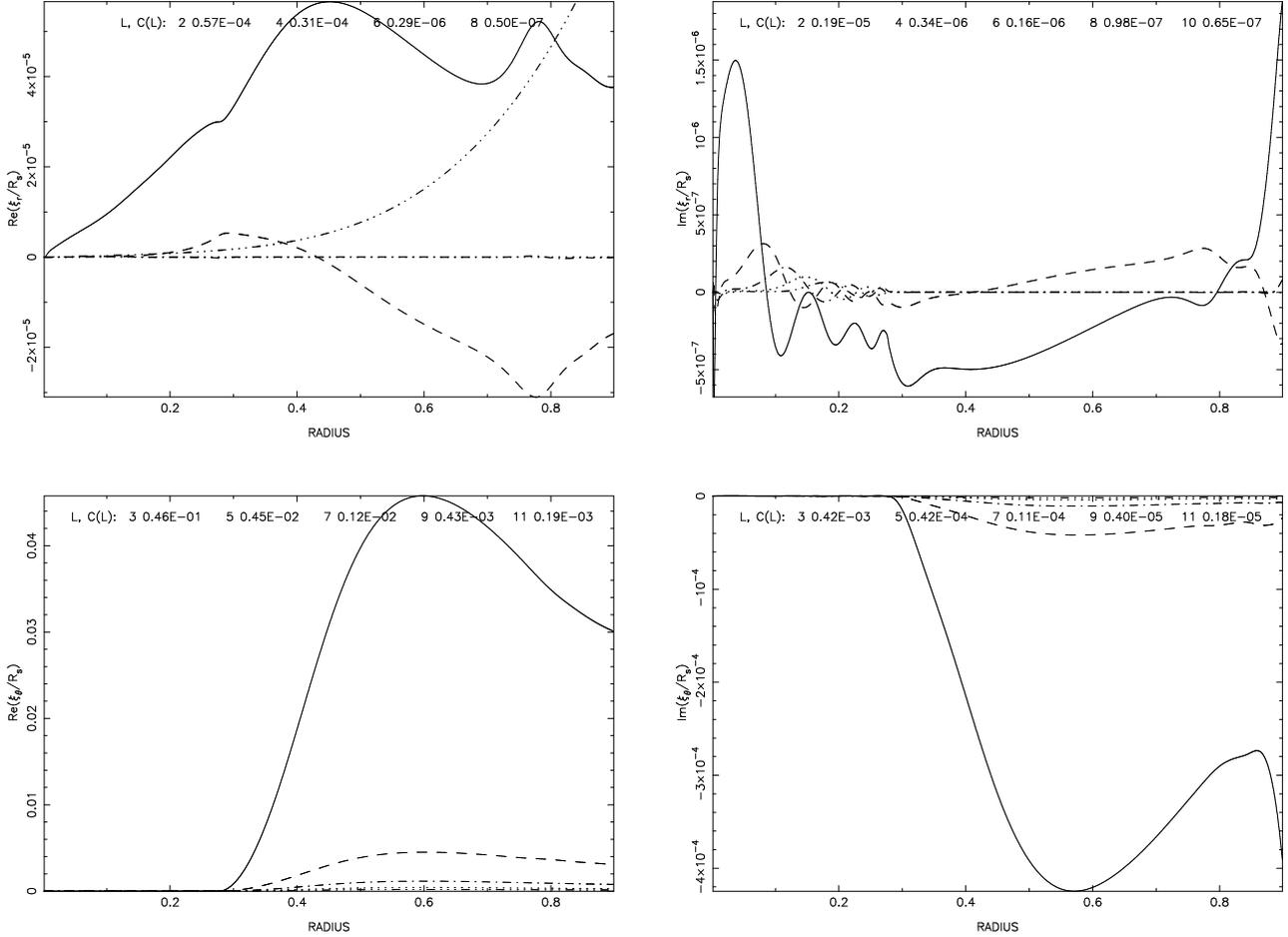}
\caption [h] {$\omgb=-0.0166$ and $\Omgs=0.1$:
corresponding to a stellar rotation period of $P_s= 5.0$ days and orbital
period $P_b\simeq 6$ days. Plotted are
the dominant Fourier-Legendre coefficients versus radius for the radial and
$\theta$-displacement for the lowest order resonance with an $r$-mode.
The various curves correspond to the consecutive $l$-values shown inside each 
panel (from left $l$-value to right $l$-value: full line, dashed line, dot-dash,
dotted and  dash-dot-dot-dot). The real number $C(l)$ given after each $l$-value
is the maximum value attained over the star for that particular FL-coefficient.
In the $\xi_r$ panel the dash-dot-dot-dot line gives the corresponding amplitude
of the equilibrium tide. The surface layers ($i>1300$) have been cut away
in order to see the smaller amplitude structure of Im($\xi_r$) in the interior.}
\label{fr1a}
\end{figure}

Figs.~\ref{fr1a} shows the radial distribution of the
dominant FL-coefficients of the  radial and $\theta$-displacement 
for the strong resonance at $\omgb=-0.0166$
with the lowest order $r$-mode that could be found for a star rotating at 
$\Omgs=0.1$. In the radiative envelope the radial displacement is 
typically three orders of magnitude smaller than the $\theta$-displacement which
shows the (quasi-~)toroidal character of this mode.

\begin{figure}
\psfig{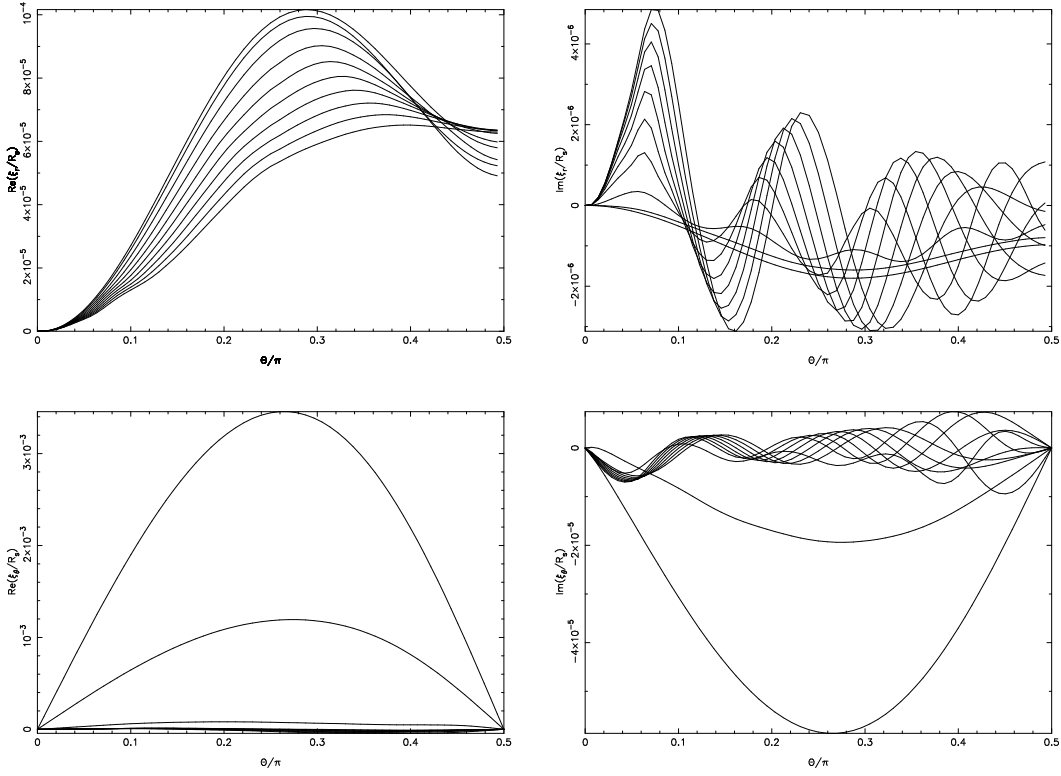}
\caption[t]{$\omgb=-0.0166$ and $\Omgs=0.1$:
the  radial and $\theta$ displacement in the outer part of the convective core 
(radial zones in the range $i=350 - 450$) as a function of $\theta$. The real 
parts show the predominant $l=2$ ($\xi_r$) and $l=3$ ($\xi_{\theta}$) dependency
on $\theta$, while the oscillatory imaginary parts indicate the excitation of
inertial waves. The shortest lengthscales were damped 
by viscosity.} 
\label{fr1b}
\end{figure}

\begin{figure}
\psfig{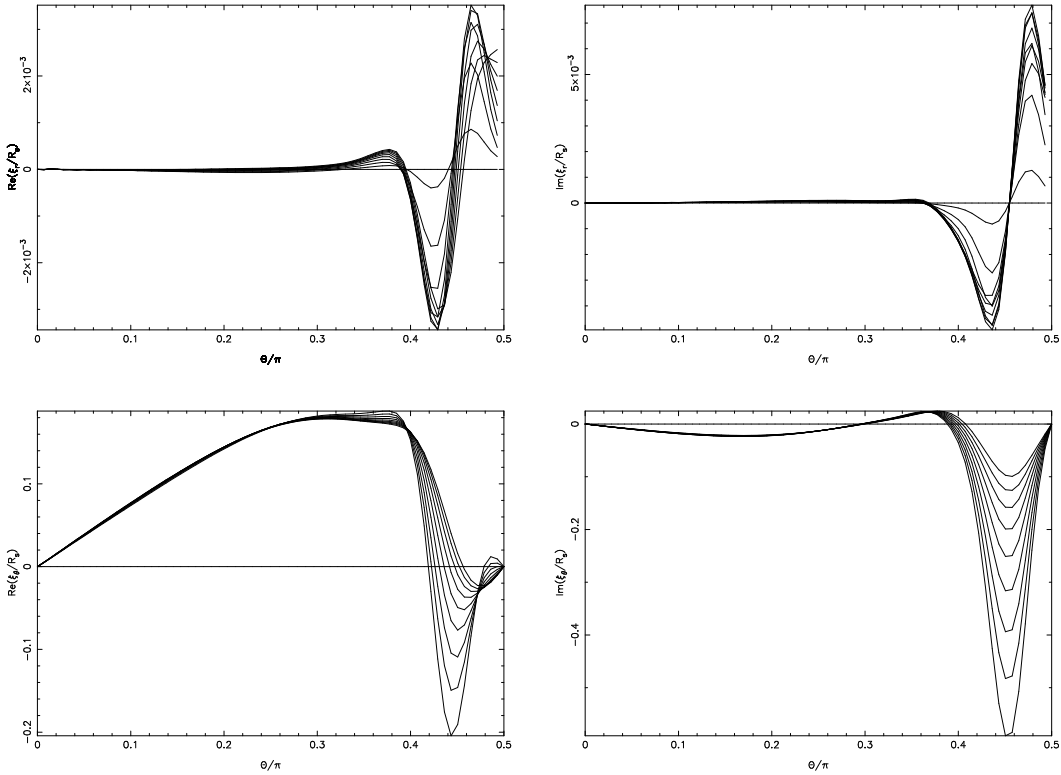}
\caption[b]{$\omgb=-0.0166$ and $\Omgs=0.1$:
the  radial and $\theta$ displacement in the region  between the
two convective surface shells ($i=1650 -1750$) as a function of $\theta$.}
\label{fr1c}
\end{figure}

For this low order $r$-mode resonance, $|\vc{\xi}|$ attains values
large compared to the equilibrium tide (Fig.~\ref{fr1a}) not only in the
radiative envelope, but also in the convective core.
The response in the core is oscillatory, both in $r$ and $\theta$
(Fig.~\ref{fr1b}) indicating the excitation of an inertial core mode.
Fig.~\ref{fr1c} shows the $\theta$-dependence of the radial and $\theta$ 
displacement in the convective shell region.
 
When $\omgb$ is made more negative than $\omgb\simeq -\frac{m \Omgs}
{l(l+1)}$ the imaginary parts of the
perturbed quantities decrease substantially. The imaginary part of $\rho^{\p}$ 
shows weak oscillations in the core and envelope, with an evanescent zone in 
between (from about i=700 to i=1100). At the boundary of the convective core 
a (relatively) strong spike occurs. When $\omgb$ is made more negative, a very 
short wavelength response appears just outside this spike, which starts to extend
outwards, grow in amplitude, and fill up the `evanescent' region.
At $\omgb=-0.0375$ (for $\Omgs=0.1$)
some 80 radial nodes occur in the radiative envelope. It appears that, 
by decreasing the forcing frequency,
a transition has been made from the r-modes to the rotationally modified,
retrograde g-modes. For more negative frequencies the number of radial nodes
decreases  and we proceed through resonances with lower and lower order g-modes,
until the limit $\omgb=-\Omgs$ is approached when $\omega\rightarrow 0$. 
\clearpage

\subsection{Tidal torque and stellar spin-up/down time}
The radiative damping in the surface layers
of tidally excited non-radial oscillation modes introduces a phase shift
between the mode and the forcing potential. This phase shift
gives rise to a net tidal torque on the secondary 

\eq {\cal T}_{\rm ext}= 2 \pi f  \int_0^{R_{\rm
s}} \int_{-1}^{1} Im (\rho^{\p})\, r^4 P^2_2(\mu) \,\, {\rm d}r\, {\rm d}\mu
\label{eqtorq} \ee

When the secondary rotates slower than synchronously ($\Omgs < \omega$) the  net
torque is positive and give rise to a spin up, while for super-synchronous
stars ($\Omgs>\omega)$ the tidal wave is retrograde in the frame co-rotating 
with the secondary and yield a negative tidal torque, leading to spin down.

\begin{figure}
\psfig{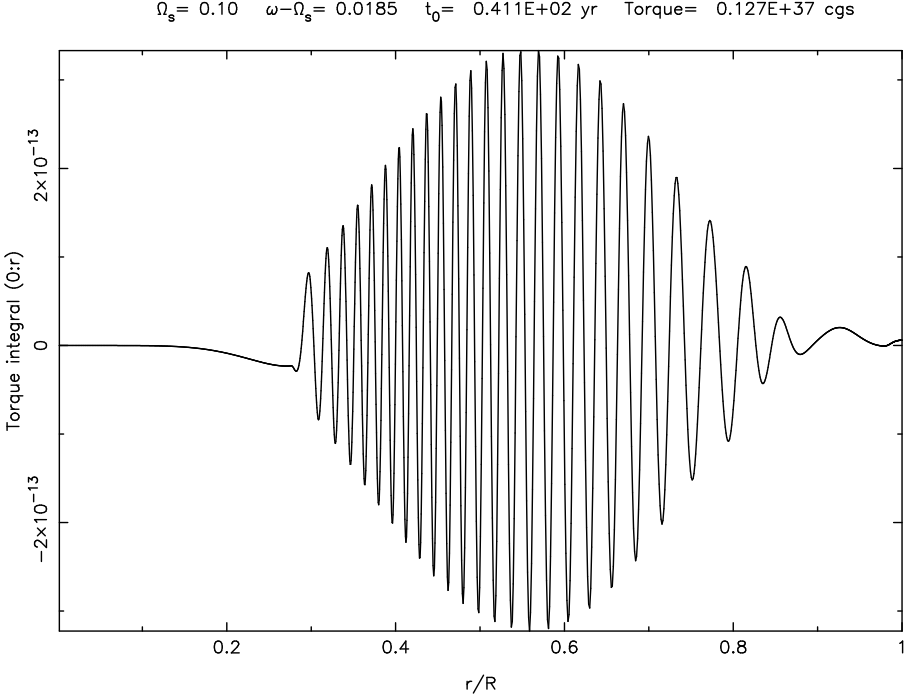}
\caption[h]
{The torque integral (\ref{eqtorq}) integrated from 0 to $r$ as
a function of $r$ for $\omgb=0.0185$ and $\Omgs=0.1$. This forcing frequency
corresponds to a strong resonance with an inertial mode in the core, causing a 
broad dip in the $t_0$ curve, see Fig.~\ref{ft001}.
The values of the plotted torque integral are given in units of $G M_s^2/R_s$.}
\label{f0185t}
\end{figure}

The secondary's spin-up/down  time-scale $t_{\rm sp}$  follows as  \[
t_{\rm sp}^{-1}=\frac{{\cal
T}_{\rm ext}}{I_{\rm s} \ol{\omega}} \]

\noi where $I_{\rm s}$ is the secondary's moment of inertia.
Let us now
define, as in our earlier studies, the `intrinsic' spin-up time $t_0$ of the
secondary as  \eq t_0(\ol{\omega})= \left(\frac{M_{\rm p}}{M_{\rm
s}}\right)^2 \left(\frac{R_{\rm s}}{D}\right)^6 t_{\rm
sp}(\ol{\omega}). \label{eqt0} \ee

\noi in which we have divided out the dependency of the $l=m=2$ forcing
on the specific binary configuration.

\begin{figure}
\psfig{figure=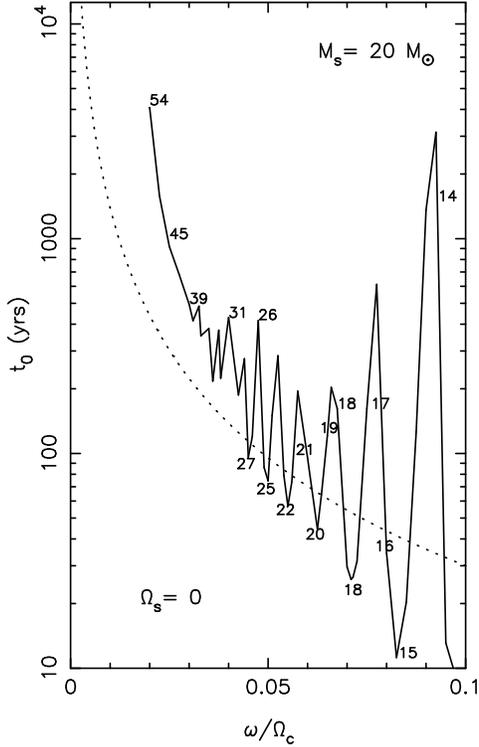,width=15cm,angle=-90}
\caption[h]{The scaled spin-up or spin-down time $t_0$ versus $\omgb$ for a 
non-rotating 20 $\Msun$ star. For comparison the low frequency limit
(\ref{eqt0asy}) is also plotted (dotted curve). The number of radial nodes is
indicated along the curve.}
\label{ft000}
\end{figure}

\noi By varying the forcing frequency $\sigb=2 \omgb$ and calculating the tidal
torque according to equation (\ref{eqtorq}) we can determine the `intrinsic' 
tidal spin-up time $t_0$ (equation \ref{eqt0}) as a function of $\omgb$.

Fig.~\ref{ft000} shows the calculated $t_0$ distribution for a non-rotating star,
whereby the numbers
along the curve indicate the number of radial nodes of the tidal oscillations.
The dotted curve corresponds to a low-frequency approximation (PS85):
\eq t_0(\omega)=\frac{k}{\Omgc} \left(\frac{M_{cc}}{M_s}\right)^{\frac{4}{3}} 
\left(\frac{R_{cc}}{\RS}\right)^{-9} \,\, \left(\frac{\omega}{\Omega_c}\right)
^{-\frac{5}{3}} \label{eqt0asy} \ee
\noi where $k$ is the star's gyration radius ($I_s=k M_s \RS^2$) and $R_{cc}$,
$M_{cc}$ are the radius and mass of the convective core. This expression is 
based on a simple model which assumes
a purely adiabatic star with a thin very non-adiabatic outer shell, so that all 
outward propagating gravity waves are absorbed by radiative damping in the outer
shell.
Expression (\ref{eqt0asy}) is similar to that obtained by Zahn (1977)
and is valid only for non-rotating unevolved  massive stars
in the limit of low forcing frequencies. In Papaloizou \& Savonije 1997 (PS97) 
the asymptotic low frequency expression for $t_0$ 
is generalized to include the effects of stellar rotation.

\begin{figure}
\psfig{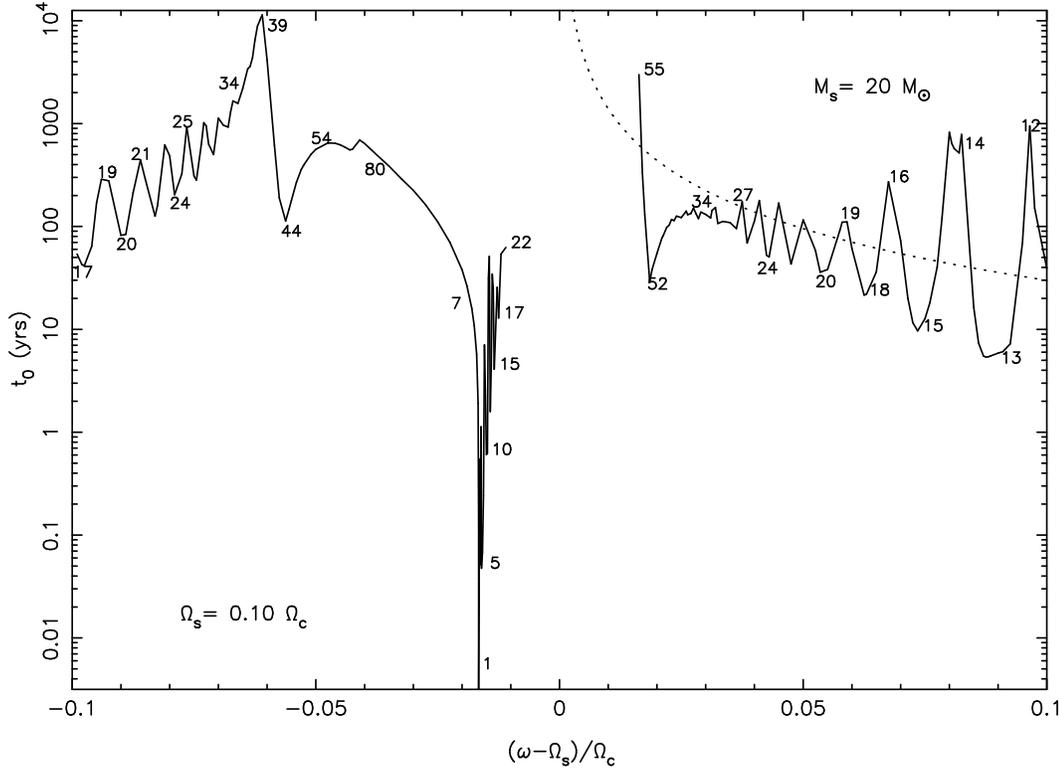}
\caption[t]{The scaled spin-up or spin-down time $t_0$ versus $\omgb$ for a 
uniformly rotating 20 $\Msun$ star at a rate $\Omgs=0.1$.
The number of radial nodes in the envelope is indicated along the curve.
For comparison the low frequency limit (\ref{eqt0asy}), whereby we have simply
replaced $\omega$ by $\omgb$,  is also plotted (dotted curve).} 

\label{ft001}
\end{figure}

\begin{figure}
\psfig{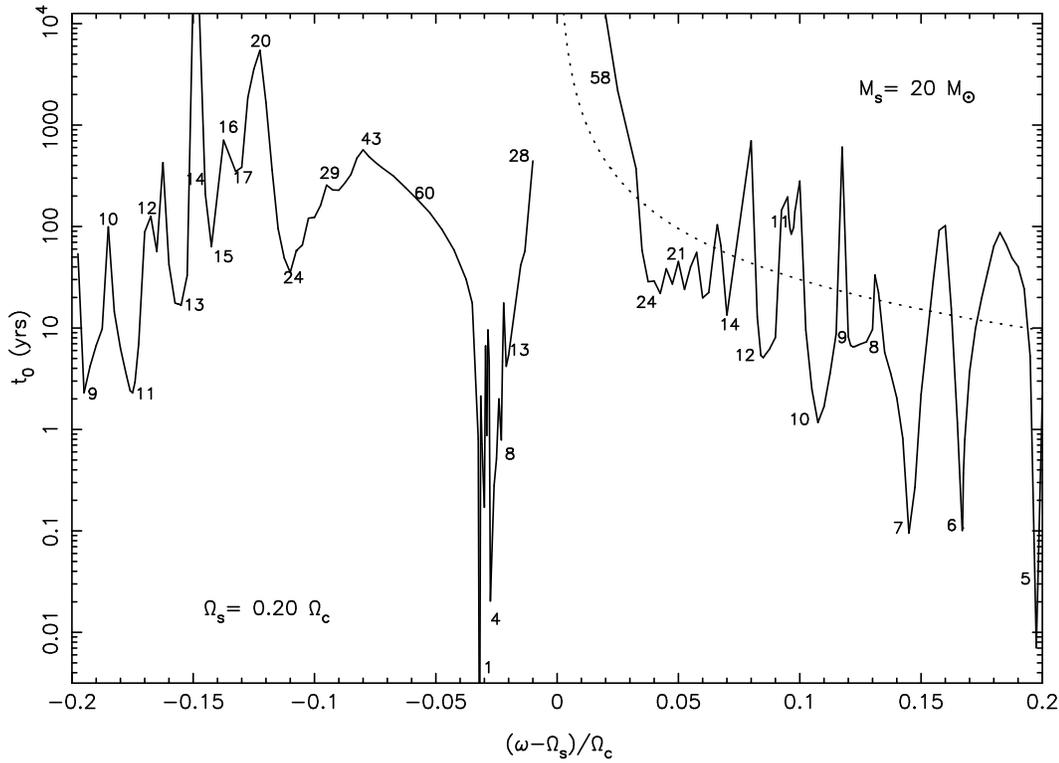}
\caption[b]{The scaled spin-up or spin-down time $t_0$ versus $\omgb$ for a 
uniformly rotating 20 $\Msun$ star at a rate $\Omgs=0.2$.
The number of radial nodes in the envelope is indicated along the curve.
For comparison the low frequency limit (\ref{eqt0asy}), whereby we have simply
replaced $\omega$ by $\omgb$,  is also plotted (dotted curve).}
\label{ft002}
\end{figure}

Figs.~\ref{ft001} and \ref{ft002} show the calculated intrinsic spin-up or down
time $t_0(\omgb)$ versus tidal frequency for a 20 $\Msun$ star rotating at a rate 
$\Omgs=0.1$ and $\Omgs=0.2$, respectively.
For low frequency g-modes, which have  many radial nodes, a strong cancellation 
effect occurs whereby the torque integral, when integrated over the envelope, 
changes sign many times.
This can be seen in Fig.~\ref{f0185t}  where the torque integral (\ref{eqtorq}),
integrated from $r=0$ to $r$, is plotted versus $r$ for  the response
at $\omgb=0.0185$. At this frequency a strong resonance occurs
with an inertial mode in the core, see next section. 
Meshpoints 1500-1800  (the convective shell region) correspond to 
$r/R > 0.98$.
Note that
for small positive values of the forcing frequency the wavelength of $g$-modes in
the envelope becomes too small ($>50$ radial nodes) for adequate resolution
on the grid.

For rotating stars which can have negative forcing frequencies when the
star spins faster than the orbital revolution of its companion, the $t_0$ curve
has an extra branch for $\omgb<0$. Rotational effects make the $t_0$ curve
qualitatively different, not only for the new branch with $\omgb<0$, but
also for $\omgb>0$. The Coriolis force couples modes with even $l-$ values, 
so that $l=2$ forcing indirectly excites $l=4,6$ .. as well. It also gives rise
to a frequency shift of the non-radial oscillations. Comparing Figs.~\ref{ft000} 
and \ref{ft001} shows that, as expected, a shift to lower eigen-frequencies has 
occurred in the rotating star (e.g. Ledoux \& Walraven 1958, SPA95). 
The response for negative forcing
frequencies is distinctly different from that for positive frequencies. 
For small negative forcing frequencies
resonances with free rotationally controlled (quasi-)toroidal r-modes show up
as strong narrow dips in the $t_0$ curve.
These strong resonances can cause a rapid spin-down of slightly
super-synchronous stars. For  more negative forcing frequencies once again
resonances with (retrograde, rotationally modified) g-modes become possible.
The more negative forcing frequencies correspond to lower order resonances
(with fewer radial nodes) that are  damped only weakly, with consequently smaller
$t_0$ values.

\clearpage

\subsection{Excitation of inertial modes in the convective core \label{sinert}}
Although the inertial modes have a dense spectrum their amplitude under tidal
forcing remains limited. This is probably partly due to coupling with envelope
modes which keeps the amplitude down because of leaking. However, the long scale
tidal force is not efficient in exciting inertial modes.
For a constant density and pressure core it can be shown (PS97)
that only two potential inertial mode resonances are excited by the
the tide, for $\ol{\omega}/\Omgs=(\pm \sqrt{105} -7)/28$. For $\Omgs=0.1$
these roots correspond to $\ol{\omega}=0.012$ and $\ol{\omega}=-0.062$.
For a core with a density and pressure gradient, like the 
one studied in this paper, inertial modes can be excited
throughout the inertial range, but generally with an amplitude smaller
than the response for the two above mentioned resonance frequencies.

\begin{figure}
\psfig{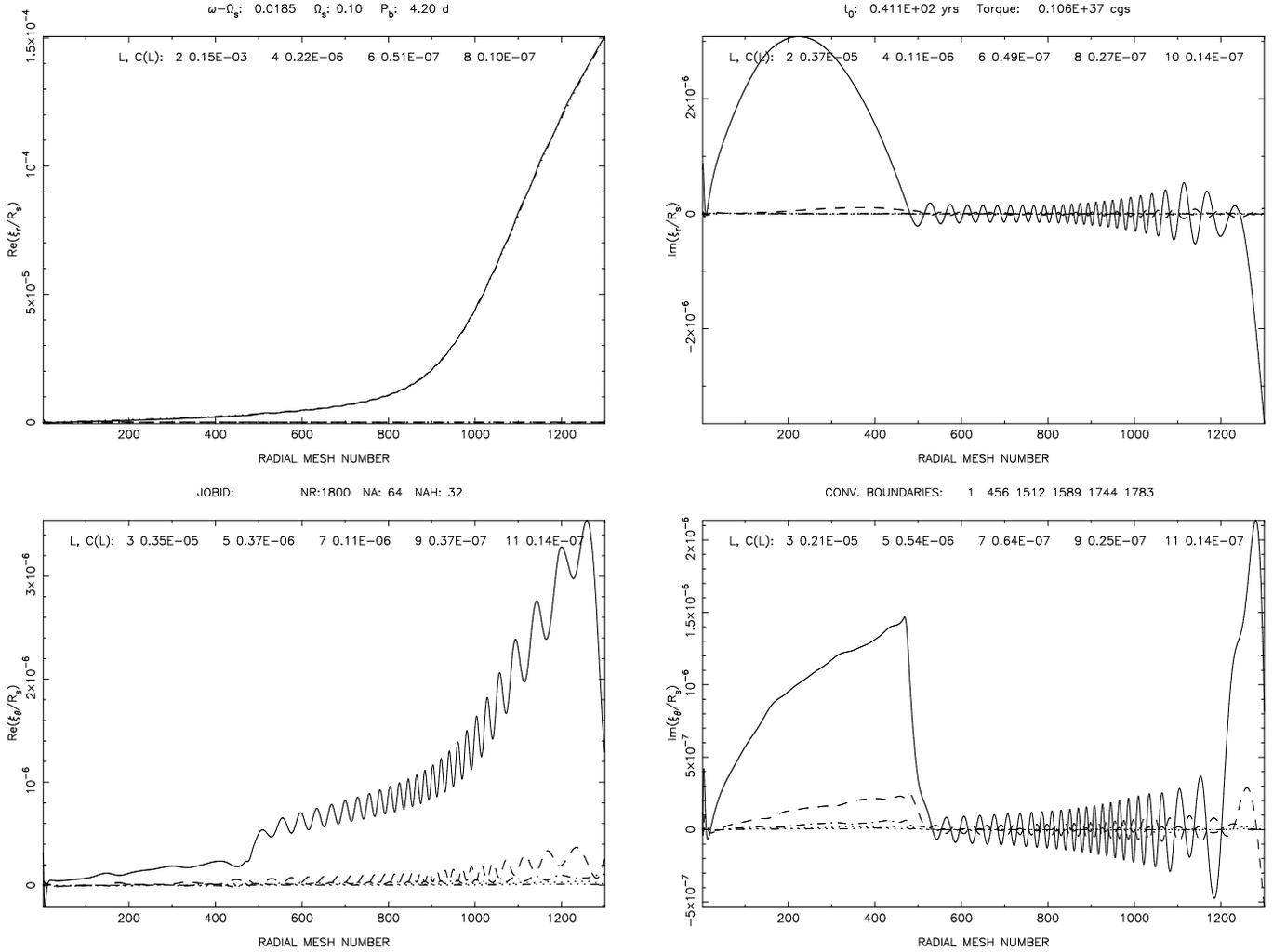}
\caption[h]{The dominant FL-coefficients for the radial and $\theta$ displacement
versus radius for the tidal response at $\omgb=0.0185$ and $\Omgs=0.1$ 
(similar to figure~\ref{fr1a}). For this
forcing frequency the tide is resonant with an inertial mode in the convective 
core. The convective core boundary is at $i\simeq 460$.
The non-adiabatic surface layer, where the amplitude becomes large,
is not shown.}
\label{f0185a}
\end{figure}

Note that in Figs.~\ref{ft001} and \ref{ft002}, which show the $t_0$ distributions
obtained by the detailed numerical calculations for the  more realistic stellar
model considered here, two broad resonance dips are visible corresponding 
roughly  with the above mentioned two resonant frequencies.
In Fig.~\ref{f0185a} we have plotted, for the positive
resonance frequency $\omgb=0.0185$ ($\Omgs=0.1$), the  FL-coefficients
of the radial and $\theta$ displacement versus radial mesh number. 
It can be seen that the imaginary parts of the resonant inertial response 
in the core attain 
amplitudes large compared to those of the (high order) g-mode in the radiative 
envelope. Apparently the inertial mode in the core excites g-modes in
the envelope to a level higher than normal, yielding the broad dip in
the $t_0$ curve.

At $\omgb\simeq 0.007$ another strong resonance occurs with an inertial mode
in the core. However, the resolution in the envelope
is too poor to determine $t_0$ for such low frequency response.

\subsection{Comparison with observations}
Giuricin, Mardirossian and Mezetti (1984 a,b,c) inspected a large sample of
early type binaries and studied their orbital eccentricity and synchronism
between orbital revolution and stellar spin rate. Their conclusion is that
Zahn's asymptotic expression (1977) for tidal effects can explain the majority of
observed eccentricities.  There remains, however, a not insignificant fraction
of relatively wide early type binaries with nearly circular orbits that
still require an explanation. Furthermore, these authors note a pronounced 
tendency
towards synchronisation up to relative radii $\RS/D\simeq 0.05$, while 
Zahn's asymptotic theory predicts a limiting separation for this corresponding to
$\RS/D\simeq 0.15$.

To synchronise a wide (circular) system with equal masses and $\RS/D=0.05$ 
during the main sequence stage of duration $\tau_{MS}$ of its early type 
components requires a time averaged `intrinsic' spin-down time $<t_0> \simeq 
(0.05)^{6}\,\,\tau_{MS}$, which for $\tau_{MS}\simeq 8 \times 10^6$ year 
corresponds to $<t_0>\simeq 0.1$ year. $\RS/D=0.10$ would require $<t_0>\simeq 10$
years. 
That demands a strong tidal response, which is difficult to attain with 
radiative damping of dynamical tides, unless the tide is in continuous resonant
interaction with a weakly damped mode.
But note that in a wide binary with $\RS/D=0.05$ and mass ratio unity 
(which corresponds to $P_b\simeq$ 30 days) an early type star rotating with
a period of 2.5 days ($\Omgs=0.2$) encounters strong
resonances with weakly damped, rotationally modified, retrograde g-modes for
which $t_0\simeq$ 1 year (Fig.~\ref{ft002}).  Hence it is expected to undergo 
significant spin down. 

\begin{figure}
\psfig{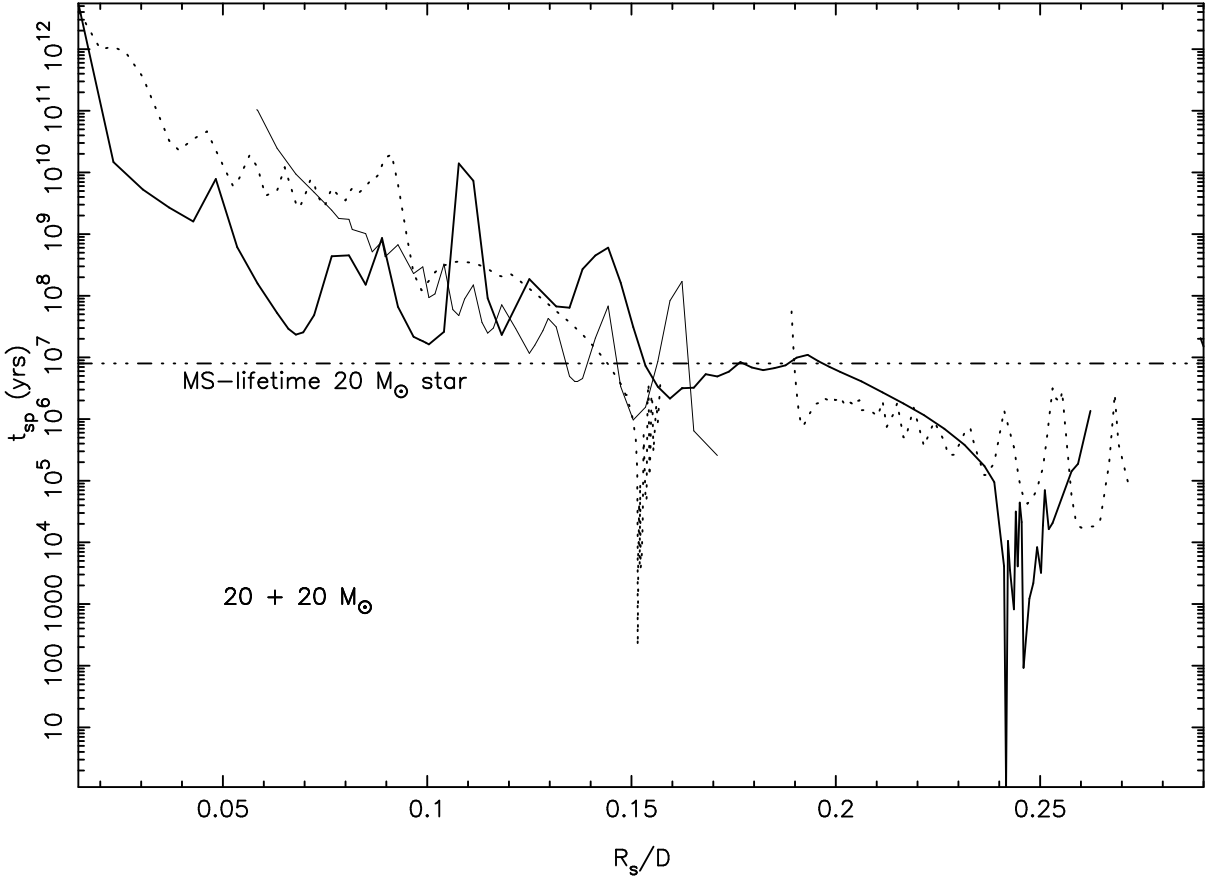}
\caption[h]{The tidal spin-up or spin-down timescale $t_{sp}$ as a function
of $\RS/D$, where $D$ is the orbital separation, for a binary system consisting 
of two 20 $\Msun$ unevolved stars.
The full fat curve corresponds to a stellar spin rate $\Omgs=0.2$,
the dotted curve to $\Omgs=0.1$ and the thin full line to a non-rotating star.
The spin up-time curve for $\Omgs=0.2$ correspond to larger values of $\RS/D$ 
than shown here.}
\label{ftsp}
\end{figure}

To visualize our numerical results we have plotted in Fig.~\ref{ftsp}
the spin-up or spin-down time as a function of $\RS/D$ for a circular binary
system consisting of two unevolved 20$\Msun$ stars, by scaling the general results
$t_0(\omgb)$ to this particular binary configuration, adopting $\Omgs=0.2$
(full fat curve), $\Omgs=0.1$ (dotted curve) and $\Omgs=0$ (thin full
line). The spin-up times for $\Omgs=0.2$ are not shown, they correspond to $\RS/D$
values larger than given in the figure.

It can be seen that with rotation the spin down time  in wide binaries
tend to be shorter, i.e. tidal effects can be significant 
even in binaries with $R/D \simeq 0.06$.
It should be taken into account that the star is subject to internal (nuclear)
evolution  causing the core to shrink and the envelope
to expand during the core hydrogen burning stage.  The envelope
expansion causes the star to spin down  by about a factor 2 during the main 
sequence stage (this factor would be larger when the envelope decouples
from the more rapidly spinning core).
The core contraction, on the other hand,  gives rise to increased gravity and
a shift of the g-mode frequencies to higher (absolute) values. Thus the internal
evolution has significant effects on the tidal evolution of the binary 
(SP84). When a mode frequency shifts to higher
frequency the radiative damping decreases and the corresponding $t_0$ goes down,
making it easier to synchronise the star. 

What could further ease synchronisation
is that the tides might induce differential
rotation whereby the surface layers get synchronised and decoupled (to some 
extent) from the more rapidly spinning interior. Then apparent synchronisation 
of a main sequence star in quite wide binaries is relatively easy. 
Goldreich \& Nicholson (1989) suggested that this is actually to be  expected
because the tidal despinning takes place where the g-modes are dissipated, i.e.
initially in the surface layers of the star. When the surface layers have been 
synchronised, the gravity waves (excited near the core boundary) can no longer
penetrate to the surface.
When  gravity waves approach the corotating surface layers their
wavelength keeps diminishing until they are fully absorbed. The absorbed negative
angular momentum carried by the gravity waves
then leads to further spin down of the layers immediately beneath the 
corotating layers. Adopting this picture, the more rapidly spinning interior
keeps its initial rotation rate with a slowly rotating, synchronized shell on top
of it. Shear instabilities or other processes of angular momentum redistribution
are expected to smooth this discontinuous angular velocity profile to some extent.
Assuming that the $t_0$ curve for the original uniform rotation rate remains
approximately applicable, we deduce from Fig.~\ref{ftsp} that for $\Omgs=~0.2$ 
the  spin down time $t_{sp}$ (required to reach corotation) is on average roughly
10 times longer than the star's main sequence lifetime for systems with $0.15 \geq
R/D \geq 0.06$. Hereby we take into account that structural changes of the star
due to stellar evolution shifts its modal spectrum, so that the star cannot remain
stuck on a bump in the $t_0$ curve.
This would correspond to synchronisation of the outer several percent of the 
stellar mass, i.e. synchronisation  down to a depth of  $r/\RS \simeq 0.6$
during the main sequence phase.
It suggests that tides in rotating stars may cause `apparent' synchronisation in
quite wide binary systems.

\section{Discussion}
We have studied the fully non-adiabatic tidal response of a slowly and
uniformly rotating  20 $\Msun$ ZAMS  star to an  $l=m=2$ perturbing
potential in the low frequency, inertial regime.
We solved the partial differential equations governing the
response  using a 2D-implicit finite difference scheme. In the work described
here, the Coriolis force was included but the centrifugal force was
neglected.  This enabled us to work with spherically symmetric equilibrium
models.

Our results show that, as expected, rotational effects can significantly
influence the tidal response. 
We have extended the study of tidal effects to super-synchronous stars in which
retrograde r-modes and rotationally modified retrograde g-modes are
excited. The r-modes give rise to efficient spin-down in slightly
super-synchronous stars, while the retrograde g-modes can give efficient
despinning of highly super-synchronous early type stars in wide binaries.
For all frequencies in the inertial regime inertial waves are excited in the core
and convective surface shells. Consistent with analytic results derived for a
simplified model with a constant density core (Papaloizou and Savonije 1997)
the detailed numerical results show the existence of frequency intervals
for which the tidal response is significantly enhanced due to resonance with an
inertial mode in the core. The calculated broad resonance features indicate that
the resonant inertial mode in the (almost adiabatic) core 
is damped by exciting gravity modes in the envelope to a level higher than normal.
Assuming that the tides induce differential rotation  by synchronising  the outer
parts of the early type star, while leaving its interior roughly unaltered,
synchronisation  may be achieved in a substantial part of the envelope
during the main sequence phase, even in quite wide binary systems. 
This could explain the observed approximate synchronous rotation in  some wide
binary systems with $R/D \simeq 0.05$.

\vspace{0.5cm}
\noi {\bf Acknowledgment}\\This work was sponsored by the Stichting Nationale
Computerfaciliteiten (National Computing Facilities Foundation, NCF) for the
use of supercomputing facilities, with financial support from the Netherlands
Organization for Scientific Research  (NWO).
We thank Marnix Witte for his help with the implementation of the opacity tables.
JCP thanks the Anton Pannekoek Institute, and GJS is grateful to QMW for 
hospitality.

\clearpage

\appendix
\section{The Finite Difference Equations (FDE's)}
We solve, as function of $r$ and $\theta$,
the set (\ref{eqmr})--(\ref{eqFt}) of seven linearized partial
differential equations, together with boundary conditions, for the
seven complex quantities $ \xi_r$, $\xi_{\theta}$, $\xi_{\varphi}$,
$\bra{\frac{\rho^{\p}}{\rho}}$, $\bra{\frac{T^{\p}}{T}}$,
$\bra{\frac{F^{\p}_r}{F}}$ and $\bra{\frac{F^{\p}_{\theta}}{F}}$
by rewriting them as a set
of finite difference equations on a polar $(r, \theta)$ grid in a meridional
plane of the star. Because the forcing has symmetry about the equatorial
plane ($\theta=\pi/2$) we need only consider the first quadrant: $0 \leq
\theta \leq \pi/2$.
The index $i=1,\Nr$ runs along radius (from $r=0$ to $r=R_{\rm s}$) and the index
$j=1,\Na$ runs in the $\theta$-direction (from $\theta=0$ to
$\theta=\pi/2$).

\begin{figure}
\psfig{figure=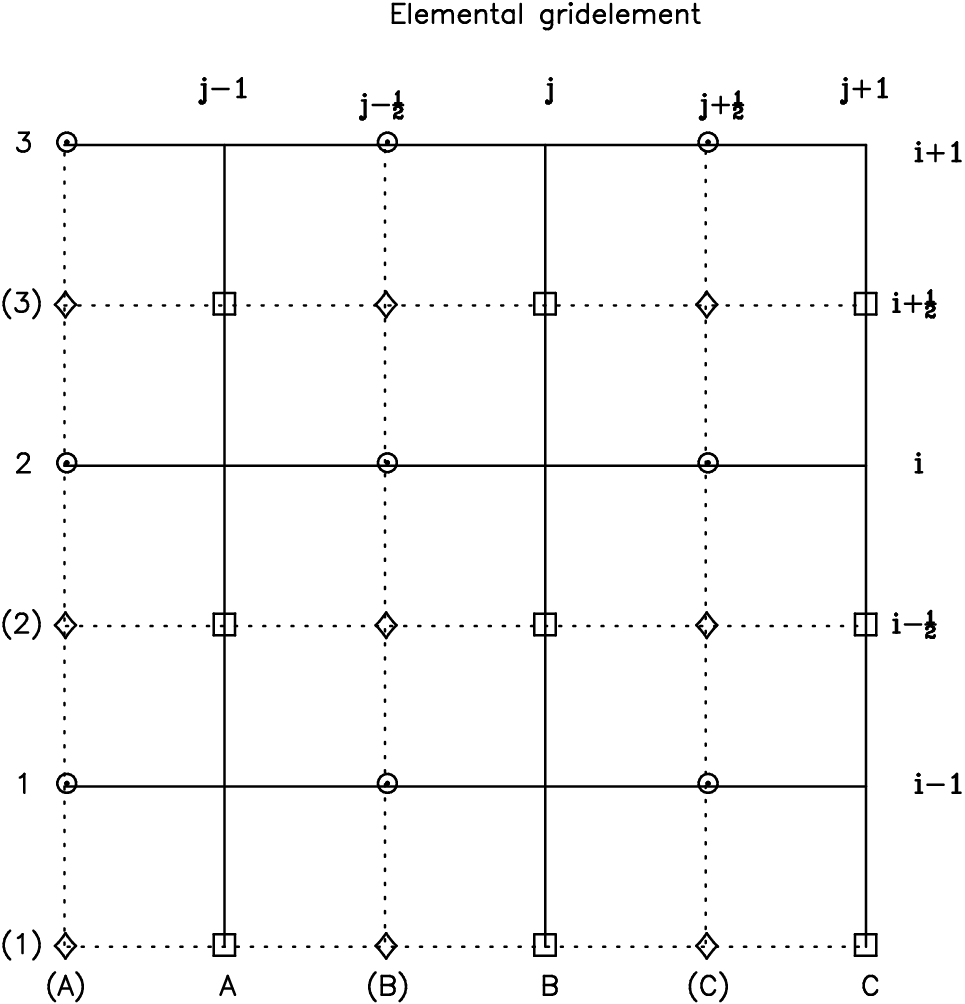,width=8cm,angle=-90}
\caption[h]{Schematic picture of the elemental gridcel for the three-level finite
difference scheme used. The lines $j=$constant all intersect at the
stellar centre. The symbols $\odot$, $\Box$ and $\Diamond$ denote the location
where the different variables are defined and the locus where the 7 difference
equations are centred.\\
$\odot$ corresponds to $\xi_r$ and $F^{\p}_r$;  centre of equations (A1,A6)\\
$\Box$ corresponds to $\xi_{\theta}$ and $F^{\p}_{\theta}$ ; centre of equations
(A2,A7) \\
$\Diamond$ corresponds to $\xi_{\varphi}$, $\rho^{\p}$ and $T^{\p}$ ;
centre of equations (A3,A4 and A5) }
\label{fgridel} \end{figure}

Note that for stable numerical results care is needed to construct the FDE's.
The set of FDE's given below yielded satisfactory numerical results. These FDE's
approximate the PDE's in such a way that for $\Omgs\rightarrow 0$ 
they too become separable in $r$ and $\theta$.
For numerical stability
we use a staggered mesh, so that the various perturbed quantities to be solved for
are not all defined at the same location.
See Fig.~\ref{fgridel} for a schematic picture of an elemental gridcel for the
three-level difference scheme used in the calculations.

The adopted radial grid is not equidistant. The $\theta$-grid is for $\theta
>\frac{\pi}{4}$ equidistant in $\mu=\cos{\theta}$ with $\Delta\mu=\mu_j-\mu_{j-1}$
 while it is for
 $\theta \le \frac{\pi}{4}$  equidistant ($\Delta \nu=\nu_j-\nu_{j-1}$) 
 in $\nu=\sin{\theta}$.
With this $\theta$-grid we have for $N_{\theta}=64$ adequate resolution, 
especially both near the rotation axis and near the stellar equator.
At the transition point $j=N_h$ (corresponding
to $\theta=\frac{\pi}{4}$)
$\nu=\mu$. However, because of the staggered mesh, for $2^{nd}$ order derivatives
expressions for a non-equidistant ($\theta$-)grid should be taken at that point.

\noi The radial equation of motion (\ref{eqmr}) is approximated by
\eqa
& \left[\sigb^2 \rho_i \RS\right] \xir_{i,\jmh}   
+\left[\ZUN \sigb \Omgs \nu_{\jmh} \rho_i \RS \right] \xip_{\imh,\jmh}   
+\left[\ZUN \sigb \Omgs \nu_{\jmh} \rho_i \RS \right] \xip_{\iph,\jmh}  \nn  \\
+ & \left[\frac{{P \chi_{\rho}}_{\imh}}{\DELH} +\h\DPRI\right]\rh_{\imh,\jmh} 
-\left[\frac{{P \chi_{\rho}}_{\iph}}{\DELH} -\h\DPRI\right] \rh_{\iph,\jmh} \nn\\
+ & \left[ \frac{{P \chi_T}_{\imh}}{\DELH} \right] \tem_{\imh,\jmh}  
-\left[\frac{{P \chi_T}_{\iph}}{\DELH} \right] \tem_{\iph,\jmh}  
=6 f \rho_i r_i \nu^2_{\jmh} 
\label{fd1}
\eea

\noi The $\theta$-equation of motion (\ref{eqmt}) is approximated by
\eqa
& \left[\sigb^2 \rho_{\imh} \RS \right] \xit_{\imh,j}   
+\ZUN \sigb \Omgs \rho_{\imh} \RS \left[ \mu_{\jph} \xip_{\imh,\jph}  +
\mu_{\jmh} \xip_{\imh,\jmh} \right]   \nn \\
&- \FUa \Delta \left[\rh_{\imh,\jph}-\rh_{\imh,\jmh} \right]  
-\FUb \Delta \left[\tem_{\imh,\jph}-\tem_{\imh,\jmh} \right]
=6 f \rho_{\imh} r_{\imh} \nu_j \mu_j 
\label{fd2}
\eea

\noi where $\Delta = \frac{\mu_j}{\delnu}$ for $j \le N_h$ and
$\Delta = \frac{-\nu_j}{\delmu}$ for $j > N_h$.

\noi The $\varphi$-equation of motion (\ref{eqmp}) becomes
\eqa
&- \ZUN \sigb \Omgs \rho_{\imh} \RS \nu_{\jmh} \left[ \xir_{i-1,\jmh}+ 
\xir_{i,\jmh} \right]  
-\ZUN \sigb \Omgs \HM{\rho} \RS \mu_{\jmh} \left[ \xit_{\imh,j-1} + \xit_{\imh,j}
\right] 
+ \sigb^2 \rho_{\imh} \RS \xip_{\imh,\jmh} \nn \\ 
& +\left[\frac{\ZUN m}{\nu_{\jmh}} \FUa\right] \rh_{\imh,\jmh}  
+\left[\frac{\ZUN m}{\nu_{\jmh}} \FUb\right] \tem_{\imh,\jmh}  
=- 3\ZUN m f r_{\imh} \rho_{\imh} \nu_{\jmh} 
\label{fd3}
\eea

\noi The equation of continuity (\ref{eqco}) becomes
\eqa
& \left[ \frac{\rho_{i} r_{i}^2 \RS}{\rho_{\imh} r^2_{\imh} \DELR} \right]
 \xir_{i,\jmh} 
-\left[ \frac{\rho_{i-1} r_{i-1}^2 \RS}{\rho_{\imh} r^2_{\imh} \DELR} \right]
 \xir_{i-1,\jmh} \nn \\
& + T_1 \xit_{\imh,j-1} + T_2  \xit_{\imh,j} 
-\left[\frac{\ZUN m \RS}{\RHM \nu_{\jmh}} \right] \xip_{\imh,\jmh}
+ \rh_{\imh,\jmh} =0
\label{fd4}
\eea

where for $j\le N_h$ the terms $T_1$ and $T_2$ are defined as \\
$ T_1=\left[ \frac{(1-2 \nu^2_{\jmh})\RS}{2 \RHM \nu_{\jmh} \mu_{\jmh}}  -
\frac{\mu^2_{\jmh}\RS}{\mu_{j-1} \RHM \delnu} \right]$ and
$ T_2=\left[ \frac{(1-2 \nu^2_{\jmh})\RS}{2 \RHM \nu_{\jmh} \mu_{\jmh}}  +
\frac{\mu^2_{\jmh}\RS}{\mu_{j} \RHM \delnu} \right]$

\noi For  $j>N_h$ these terms are given by \\
$ T_1=\left[\frac{\mu_{\jmh}}{\RHM \nu_{\jmh}} \RS + \frac{\nu^2_{\jmh}}{\nu_{j-1}
\RHM \delmu} \RS \right] $ and  $
T_2=\left[\frac{\mu_{\jmh}}{\RHM \nu_{\jmh}} \RS - \frac{\nu^2_{\jmh}}{\nu_{j}
\RHM \delmu} \RS \right] $

\noi The energy equation (\ref{eqen}) is approximated by
\eqa
& \left[\DPH -\HM{\Gamma_1} \DROH\right] \frac{\RS}{2} \xir_{i,\jmh}
+\left[\DPH -\HM{\Gamma_1} \DROH\right] \frac{\RS}{2} \xir_{i-1,\jmh} \nn \\
& +\left[-\frac{\HM{\ZH}}{\DELR}\left(\frac{r^2_{i}}{\RHM^2 }+\frac{(F_i-F_{i-1})}
{2 F_{\imh}}
\right)\right] \fr_{i,\jmh} 
+\left[-\frac{\HM{\ZH}}{\DELR}\left(-\frac{r^2_{i-1}}{\RHM^2}+\frac{(F_i-F_{i-1})}{2 F_{\imh}}
\right)\right] \fr_{i-1,\jmh} \nn \\
& +T_1 \ft_{\imh,j} +T_2  \ft_{\imh,j-1}  
+\left[\chi_T+\left(\frac{m}{\RHM \nu_{\jmh}}\right)^2 \frac{\HM{\ZH}}{\DTH} 
+\right] \tem_{\imh,\jmh} 
=\left[\Gamma_1-\chi_{\rho}\right]_{\imh} \rh_{\imh,\jmh}
\label{fd5a}
\eea

where  for $j\le N_h$ \\
$ T_1=-\frac{\HM{\ZH}}{\RHM \mu_{j}}\left[\mu_{\jmh}^2 \left(\frac{1}
{\delnu}+\frac{1}{2 \nu_{\jmh}}-\h \nu_{\jmh}\right)\right]    $
and
$T_2=-\frac{\HM{\ZH}}{\RHM \mu_{j-1}}\left[\mu_{\jmh}^2 \left(\frac{-1}
{\delnu}+\frac{1}{2 \nu_{\jmh}}-\h\nu_{\jmh}\right)\right]  \nn  \\$
whereas for $j>N_h$ these terms are:\\
$ T_1=-\frac{\HM{\ZH}}{\RHM \nu_{j}}\left[\mu_{\jmh}-\frac{\nu_{\jmh}^2}{\delmu}
\right] $
and
$ T_2=-\frac{\HM{\ZH}}{\RHM \nu_{j-1}}\left[\mu_{\jmh}+\frac{\nu_{\jmh}^2}{\delmu}
\right] $

\noi The radial energy diffusion equation (\ref{eqFr}) is approximated by
\eqa
\fr_{i,\jmh} +\h\left(\left(\kappa_{\rho}\right)_i+1\right) \left[
\rh_{\imh,\jmh}+\rh_{\iph,\jmh} \right] 
+\left[ \h\left(\kappa_T \right)_i-2 -\frac{T_i}{T_{\iph}-T_{\imh}} \right]
\tem_{\iph,\jmh} \nn \\
+\left[ \h\left(\kappa_T \right)_i-2 +\frac{T_i}{T_{\iph}-T_{\imh}} \right]
\tem_{\imh,\jmh}=0 
\label{fd6}
\eea

\noi Finally, the energy diffusion in the $\theta-$direction (\ref{eqFt}) is
approximated by
\eqa
\frac{r_i-r_{i-1}}{T_i-T_{i-1}}\frac{T_{\imh}}{r_{\imh}} \Delta
\left[\tem_{\imh,\jmh}-\tem_{\imh,\jph} \right]
+\ft_{\imh,j} =0
\label{fd7}
\eea
\noi where $\Delta$ was defined in equation (\ref{fd2}).

\subsection{The viscous terms}
The viscous terms (\ref{eqvisr}-\ref{eqvist}) added to the equations
of motion contain second order derivatives and must therefore be
defined on three levels of the grid. Again the finite differences are constructed
in such a way that the expressions retain separability
in $r$ and $\theta$. To that end the derivatives with respect to $\mu$
in equations (\ref{eqvisr}-\ref{eqvist}) are for $j\le N_h$ expressed as 
\[ \pdrv{\xi_r}{\mu}=-\frac{\mu}{\nu} \ol{\xi_r} -\pdrv{\ol{\xi_r}}{\nu} \]
\noi and
\[ \pdrv{^2}{\mu^2}\xi_r= -\frac{\ol{\xi_r}}{\nu^3} + \frac{1 -2 \nu^2}{\nu^2}
\pdrv{\ol{\xi_r}}{\nu} + \frac{\mu^2}{\nu}\pdrv{^2}{\nu^2} \ol{\xi_r} \]
\noi where for $\Omgs=0$ we have $\ol{\xi_r}\equiv \frac{\xi_r}{\nu}\propto \nu $.
For $\xi_{\theta}$ the $\mu$-derivatives are expressed as
\[ \pdrv{^2}{\mu^2} \xi_{\theta}=-\left(3 \frac{\mu}{\nu} + \frac{\mu^3}{\nu^3}
\right) \drv{\ol{\xi_{\theta}}}{\nu} + \frac{\mu^3}{\nu^2} \pdrv{^2}{\nu^2} 
\ol{\xi_{\theta}} \]
where for $\Omgs=0$ we have 
$\ol{\xi_{\theta}}\equiv \frac{\xi_{\theta}}{\mu} \propto\nu$.
With $\alpha_i=\ZUN \sigb \RS (\zeta/r^2)_i$, the  radial 
derivatives to be added to the equation of radial motion (\ref{fd1}) become
\eqa
-\frac{\alpha_i}{\Delta_3} \left[ \frac{\left(\rho \zeta r^2 \right)_{\iph}
-\left(\rho \zeta r^2 \right)_{\imh}}{\zeta_i \DELH} \frac{\Delta_1}{\Delta_2}
-\frac{2 \left(\rho r^2 \right)_i}{\Delta_2} \right]
\xir_{i-1,\jmh} 
+\frac{\alpha_i}{\Delta_3} \left[ \frac{\left(\rho \zeta r^2 \right)_{\iph}
-\left(\rho \zeta r^2 \right)_{\imh}}{\zeta_i \DELH} 
\left(\frac{\Delta_1}{\Delta_2}-\frac{\Delta_2}{\Delta_1}\right)\right]
\xir_{i,\jmh} \nn \\
-\frac{\alpha_i}{\Delta_3}\left[2 \left(\rho  r^2 \right)_i 
\left(\frac{1}{\Delta_1}+\frac{1}{\Delta_2}\right) \right]
\xir_{i,\jmh} 
+\frac{\alpha_i}{\Delta_3} \left[ \frac{\left(\rho \zeta r^2 \right)_{\iph}
-\left(\rho \zeta r^2 \right)_{\imh}}{\zeta_i \DELH} \frac{\Delta_2}{\Delta_1}
+\frac{2 \left(\rho r^2 \right)_i}{\Delta_1} \right]
\xir_{i+1,\jmh} 
\eea
\noi where $\Delta_1=r_{i+1}-r_i$,
$\Delta_2=r_i-r_{i-1}$ and $\Delta_3=r_{i+1}-r_{i-1}$.

\noi The $\theta$-derivatives to be added to the equation of radial motion become
for $j \le N_h$
\eqa
\frac{\alpha_i \rho_i}{\Delta_3} \left[-\frac{3-4 \nu^2_{\jmh}}{\nu_{j-\hh}}
\frac{\Delta_1}{\Delta_2}+\frac{2 \mu^2_{\jmh}\nu_{\jmh}}{\nu_{j-\hh} \Delta_2} 
\right] \xi_{i,j-\hh} 
-\frac{\alpha_i \rho_i}{\Delta_3} \left[ 2+\frac{3}{\nu_{\jmh}^2}+2 \mu_{\jmh}^2 
\left(\frac{1}{\Delta_1}+\frac{1}{\Delta_2}\right) \right] \xir_{i,\jmh} \nn \\
-\frac{\alpha_i \rho_i}{\Delta_3} \left[\frac{3 -4 \nu^2_{\jmh}}{\nu_{\jmh}}
\left(\frac{\Delta_2}{\Delta_1}-\frac{\Delta_1}{\Delta_2} \right) \right]
\xir_{i,\jmh} 
\frac{\alpha_i \rho_i}{\Delta_3}\left[\frac{3 -4 \nu^2_{\jmh}}{\nu_{\jph}}
\frac{\Delta_2}{\Delta_1}+\frac{2 \mu^2_{\jmh}\nu_{\jmh}}{\nu_{\jph}\Delta_1}
\right] \xir_{i,\jph}
\eea
\noi where
$\Delta_1=\nu_{\jph}-\nu_{\jmh}$, $\Delta_2=\nu_{\jmh}-\nu_{j-\hh}$
and $\Delta_3=\nu_{\jph}-\nu_{j-\hh}$. Note that at the transition point $j=N_h$
the grid is not equidistant (neither in $\mu$ nor $\nu$) and that it is essential
to use the expressions for non-equidistant mesh given here.
For $j>N_h$ these derivatives become

\eqa
\frac{\alpha_i \rho_i}{\Delta_3}\left[\frac{2 \nu^2_{\jmh}}{\Delta_2}
+2 \mu_{\jmh}\frac{\Delta_1}{\Delta_2} \right]
\xir_{i,j-\hh} 
-\frac{\alpha_i \rho_i}{\Delta_3}\left[2\nu^2_{\jmh}
\left(\frac{1}{\Delta_1}+\frac{1}{\Delta_2}\right)+\frac{4}{\nu^2_{\jmh}} \right] \xir_{i,\jmh} \nn \\
+\frac{\alpha_i \rho_i}{\Delta_3}\left[2\mu_{\jmh}
\left(\frac{\Delta_2}{\Delta_1}-\frac{\Delta_1}{\Delta_2} \right) \right]
\xir_{i,\jmh} 
+\frac{\alpha_i \rho_i}{\Delta_3}\left[\frac{2 \nu^2_{\jmh}}{\Delta_1}
-2 \mu_{\jmh}\frac{\Delta_2}{\Delta_1} \right]
\xir_{i,\jph} 
\eea
\noi where now $\Delta_1=\mu_{\jph}-\mu_{\jmh}$, $\Delta_2=\mu_{\jmh}-\mu_{j-\hh}$
and $\Delta_3=\mu_{\jph}-\mu_{j-\hh}$
\noi and similarly for the $\theta$-equation of motion.

\section{Numerical solution technique}
Let us introduce the following definition for the perturbation vector
$\q{X}_{i,j}$ at each
gridpoint centre $(i,j)$, consistent with Fig.~\ref{fgridel}

\eqa
 \q{X}_{i,j} = ( \left[\frac{\xi_r}{R_{\rm s}} \right]_{i,\jmh},
 \left[\frac{F^{\p}_r}{F}\right]_{i,\jmh},
 \left[\frac{\xi_{\theta}}{R_{\rm s}}\right]_{\imh,j}, 
 \left[\frac{F^{\p}_{\theta}}{F}\right]_{\imh,j}, 
 \left[\frac{\xi_{\varphi}}{R_{\rm s}} \right]_{\imh,\jmh},
 \left[\frac{\rho^{\p}}{\rho}\right]_{\imh,\jmh},
 \left[\frac{T^{\p}}{T}\right]_{\imh,\jmh} )^t 
\eea

\noi   where $F$, $\rho$ and $T$ respectively denote the local
unperturbed energy-flux, density and temperature and $\RS$ the stellar radius.

Then the finite difference equations (\ref{fd1}-\ref{fd7})
corresponding to the 7 linearized  partial differential equations
(\ref{eqmr})$-$(\ref{eqFt}) extended with the viscous terms
can for each step ($i,j)$ be expressed as a single matrix equation

\eqa
A_1 \q{X}_{i-1,j-1} +  B_1 \q{X}_{i-1,j}  +  C_1 \q{X}_{i-1,j+1} + 
A_2 \q{X}_{i,j-1}   +  B_2 \q{X}_{i,j}    +  C_2 \q{X}_{i,j+1}   + 
A_3 \q{X}_{i+1,j-1} +  B_3 \q{X}_{i+1,j}  +  C_3 \q{X}_{i+1,j+1} + D= 0 
\label{eqfd}  \eea

\noi where the $7\times 7$-matrices $A_l$, $B_l$ and $C_l$, for $l=1,2,3$ are 
complex and defined in terms of the unperturbed 
stellar model.

The elimination procedure that solves for the perturbation vectors
$\q{X}_{i,j}$ is initiated by applying the boundary conditions at the
stellar centre ($i-1=0$), where we  require $\xi_{r}$ and $F^{\p}_r$ to vanish.
The remaining five components of the perturbation vector $\q{X}_{0,j}$ are located
outside the grid, see Fig.~(\ref{fgridel}), and need not be defined.
Hence, for $i=1$ the terms with the coefficient matrices $A_1$, $B_1$
and $C_1$ disappear from the finite difference equation (FDE) (\ref{eqfd}) 
for all $j$. 

By applying the boundary conditions at the stellar rotation axis ($j-1=0$)
and requiring $\xi_{\theta}$ and $F^{\p}_{\theta}$ to vanish, and noting that the 
remaining five components of the perturbation vectors $\q{X}_{i,0}$ need not be
defined (see Fig.~\ref{fgridel}),
we conclude  that for $j=1$ the terms with the coefficient matrices $A_l$, with 
$l=1,2,3$, also disappear from the FDE (\ref{eqfd}) for
all $i$. 

For $i=1$, $j=1$ the simplified FDE's can thus be written as
\eq B_2 \q{X}_{1,1}+ C_2 \q{X}_{1,2}+ B_3 \q{X}_{2,1} + C_3 \q{X}_{2,2}+D=0 \ee
By inverting matrix $B_2$ we immediately find the relation:
\eq \q{X}_{1,1}=-B_2^{-1} C_2 \q{X}_{1,2}- B_2^{-1} B_3 \q{X}_{2,1}-
B_2^{-1} C_3\q{X}_{2,2}-B_2^{-1}D \label{eqrl0} \ee

The elimination procedure is continued by stepping up j for  given $i=1$.
For $j=2$ the vector $\q{X}_{1,j-1}$ no longer vanishes, but it can be eliminated
by substituting relation (\ref{eqrl0}) for 
$\q{X}_{1,j-1}$ into  the FDE. After again multiplying by the inverted
matrix $B_2^{-1}$ we obtain  a relation of the form

\eq\q{X}_{1,j}=\alpha^{1,j}\q{X}_{1,j+1}+\sum_{l=1}^{j+1}\beta^{1,j}_l\q{X}_{2,l} 
 + \gamma^{1,j} \label{eqrl1}\ee

where $\alpha^{1,j}$, $\beta^{1,j}_l$ are complex $7 \times 7$-matrices and
$\gamma^{1,j}$ is a 7-vector.

\noi For $j>2$ expression (\ref{eqrl1}) is applied after every step $j$ to 
eliminate $\q{X}_{1,j}$ from all previous steps ($j-1$,$j-2$, ... 1).
After repeating the elimination procedure until $j=\Na$ we can apply
the equatorial symmetry conditions

\eq\q{X}_{1,\Na+1}={\cal{S}} \q{X}_{1,\Na} \ee

\noi where $\cal{S}$ is a diagonal $7 \times 7$-matrix. Its diagonal elements 
corresponding to $\xi_{\theta}$ and $F^{\p}_{\theta}$ are $-1$, while all 
remaining diagonal elements are 1.
Using these symmetry relations to eliminate $\q{X}_{1,\Na+1}$,
relation (\ref{eqrl1}) for $\q{X}_{1,\Na}$ can be re-expressed as

\eq\q{X}_{1,\Na}=\sum_{l=1}^{\Na}\beta^{1,\Na}_l \q{X}_{2,l} + \gamma^{1,\Na} \ee
where $\beta^{1,\Na}_l$ and $\gamma^{1,\Na}$ have been redefined.

After substituting the above expression for $\q{X}_{1,\Na}$ in all previously 
obtained results, we have eliminated all unknowns $\q{X}_{1,k}$ on the right hand
side of expression (\ref{eqrl1})
and we end up with the set of relations (for $k=$1,2,3, .. $\Na$)

\eq \q{X}_{1,k}=\sum_{l=1}^{\Na}\beta^{1,k}_l \q{X}_{2,l}+\gamma^{1,k}
 \label{eqrl2}\ee
\noi where the $\beta^{1,k}_l$ and $\gamma^{1,k}$ have been updated.

We can now step up $i$ and start a new
strike $j=1,2, ... \Na$ (for $i=2,3,...\Nr$). For $i>1$ the perturbation 
vectors $\q{X}_{i-1,k}$ no longer
vanish in the FDE. However, relations (\ref{eqrl2}) obtained in the
previous strike $i-1$ can now be used to eliminate these perturbation vectors  
from
the FDE. To this end the latter relations must be updated after every elimination
step $(i,j)$ to eliminate all $\q{X}_{i,l}$ with $l<j$ from the summation on the
right hand side of (\ref{eqrl2}). For step $(i,j)$ this results in an expression
of the form (for $k=j-1, j, j+1, .... \Na$)

\eq\q{X}_{i-1,k}=\alpha^{i-1,k}\q{X}_{i,j}+\sum_{l=1}^j\beta^{i-1,k}_l\q{X}_{i+1,l}
+\sum _{l=j+1}^{\Na}\beta^{i-1,k}_l \q{X}_{i,l} + \gamma^{i-1,k}\label{eqrl3}\ee

From the previous elimination step $(i, j-1$) of the current strike $i$ we have
\eq\q{X}_{i,j-1}=\alpha^{i,j-1}\q{X}_{i,j}+\sum_{l=1}^j\beta^{i,j-1}_l\q{X}_{i+1,l}
+\sum _{l=j+1}^{\Na} \beta^{i,j-1}_l\q{X}_{i,l} + \gamma^{i,j-1}\label{eqrl4}\ee

\noi After substituting these results in the FDE (\ref{eqfd})  for step $(i,j)$, 
all terms left of $B_2 \q{X}_{i,j} + ...$ can be eliminated and the FDE is
cast in the form
\eq B_2\q{X}_{i,j}+C_2\q{X}_{i,j+1}+\sum_{l=1}^{j+1}\beta^{i,j}_l\q{X}_{i+1,l} +
\sum _{l=j+2}^{\Na} \beta^{i,j}_l \q{X}_{i,l} + D =0 \label{eqfd1} \ee
\noi where $B_2$, $C_2$ and $D$ have been redefined to collect the various
coefficients of the corresponding vectors.

By again multiplying the FDE with the
inverted matrix $B_2^{-1}$ the vector $\q{X}_{i,j}$ can be expressed as

\eq\q{X}_{i,j}=\alpha^{i,j}\q{X}_{i,j+1}+\sum_{l=1}^{j+1}\beta^{i,j}_l\q{X}_{i+1,l}+
 \sum_{l=j+2}^{\Na} \beta^{i,j}_l \q{X}_{i,l} + \gamma^{i,j} \label{eqrl5} \ee

\noi which is of the same format as the result of the previous elimination step
 (\ref{eqrl4}).
The current elimination step $(i,j)$ is terminated by
updating expressions (\ref{eqrl3}-\ref{eqrl4}). To this end 
expression (\ref{eqrl5}) for $\q{X}_{i,j}$ is substituted on their right hand 
sides.
Equation (\ref{eqrl3}) must be updated for all $k \ge j$.

After repeating this elimination procedure until $j=\Na$ the equatorial boundary 
conditions can be applied  after which the stored relations for this strike
$i$ can be written as ($k=1,2, ... \Na$)
\eq\q{X}_{i,k}=\sum_{l=1}^{\Na} \beta^{i,k}_l\q{X}_{i+1,l}+\gamma^{i,k} \label{eqrl6}
\ee
\noi with which we can start a new strike $i+1$ up to $i=\Nr$, which corresponds
to the stellar surface. At the surface we require the Lagrangian pressure
perturbation $\delta P$ to vanish and the radial flux perturbation to fulfill
Stefan-Boltzmann's relation. In this way  $\q{X}_{\Nr+1,J}$ can
be ignored, so that the first summation on the right hand side of relations 
(\ref{eqrl3}, \ref{eqrl4}) and equation(\ref{eqfd1}) vanishes. After working 
through the last strike 
$i=\Nr$ we finally arrive at $j=\Na$. Note that now also the second  summation
in the above relations has vanished.
Substituting the thus simplified relations (\ref{eqrl3}-\ref{eqrl4}) and applying
the equatorial symmetry
relations, all terms in the FDE can now be collected in $B_2$ and $D$ only
\eq B_2 \q{X}_{\Nr,\Na} + D =0 \ee
Solving this equation we find the first perturbation vector $\q{X}_{\Nr,\Na}=
-B_2^{-1} D $. We can find all other perturbation vectors by stepping back
through the entire elimination procedure. E.g.  having solved ${\q{X}_{\Nr,\Na},
\q{X}_{\Nr,\Na-1}, ...,\q{X}_{\Nr,j+1}}$, we find $\q{X}_{\Nr,j}$ by direct 
substitution into the stored relations (\ref{eqrl5})
\eq\q{X}_{\Nr,j}=\alpha^{\Nr,j}\q{X}_{\Nr,j+1}+\sum_{l=j+2}^{\Na}\beta^{\Nr,j}_l
\q{X}_{\Nr,l} + \gamma^{\Nr,j} \ee
\noi Having solved in this way for all $\q{X}_{\Nr,l}$, we find $\q{X}_{\Nr-1,l}$
from relations (\ref{eqrl6}), etc.

\end{document}